\documentclass[journal,12pt,onecolumn,draftclsnofoot]{IEEEtran}

\IEEEoverridecommandlockouts
\usepackage[utf8]{inputenc}
\usepackage{bm}
\usepackage{amsfonts}
\usepackage{amssymb}
\usepackage{amsmath}
\usepackage{algorithm}
\usepackage{algpseudocode}
\usepackage{graphicx}
\usepackage{siunitx}
\usepackage{cite}
\usepackage[font=small]{caption}
\usepackage{multicol}

\begin{document}

\title{Approximations of the Aggregated Interference Statistics for Outage Analysis in Massive MTC}

\author{Sergi~Liesegang,~\IEEEmembership{Student Member,~IEEE}, Antonio~Pascual-Iserte,~\IEEEmembership{Senior Member,~IEEE}, Olga~Mu\~noz,~\IEEEmembership{Member,~IEEE}
\thanks{This manuscript is an extended version of the conference paper: S. Liesegang, O. Mu{\~{n}}oz, and A. Pascual-Iserte, ``Interference Statistics Approximations for Data Rate Analysis in Uplink Massive MTC,'' in proceedings of the \textit{2018 IEEE Global Conference on Signal and Information Processing}, Anaheim, CA, USA, November 2018, pages 176-180.}
\thanks{The work presented in this paper was carried out within the framework of the project 5G\&B RUNNER-UPC (TEC2016-77148-C2-1-R (AEI/FEDER, UE)), the research network RED2018-102668-T Red COMONSENS and the~FPI grant BES-2017-079994, funded by the Spanish Ministry of Science, Innovation and Universities; and~the grant 2017 SGR 578, funded by the Catalan Government (AGAUR, Secretaria d'Universitats i Recerca, Departament d'Empresa i Coneixement, Generalitat de Catalunya).}
\thanks{The authors are with the Department of Signal Theory and Communications,
Universitat Polit\`ecnica de Catalunya, 08034 Barcelona, Spain (e-mails: sergi.liesegang@upc.edu, antonio.pascual@upc.edu, olga.munoz@upc.edu).

DOI: 10.3390/s19245448

}}

\markboth{Accepted paper at MDPI Sensors (Sensors 2019, 19, 5448)}
{}

\maketitle

\begin{abstract}
This paper presents several analytic closed-form approximations of the aggregated interference statistics within the framework of uplink massive machine-type communications (mMTC), taking into account the random activity of the sensors. Given its discrete nature and the large number of devices involved, a continuous approximation based on the Gram--Charlier series expansion of a truncated Gaussian kernel is proposed. We use this approximation to derive an analytic closed-form expression for the outage probability, corresponding to the event of the signal-to-interference-and-noise ratio being below a detection threshold. This metric is useful since it can be used for evaluating the performance of mMTC systems. We analyze, as an illustrative application of the previous approximation, a scenario with several multi-antenna collector nodes, each equipped with a set of predefined spatial beams. We consider two setups, namely single- and multiple-resource, in reference to the number of resources that are allocated to each beam. A graph-based approach that minimizes the average outage probability, and that is based on the statistics approximation, is used as allocation strategy. Finally, we describe an access protocol where the resource identifiers are broadcast (distributed) through the beams. Numerical simulations prove the accuracy of the approximations and the benefits of the allocation strategy.
\end{abstract}


\section{Introduction} \label{sec:1}
Machine-type communications (MTC) have drawn a lot of attention in the past years among academic and industrial communities. They can be defined as a set of transmissions between connected terminals with no human interaction~\cite{Wan17}, which will enable the creation of a myriad of applications such as the Internet-of-Things (IoT) \cite{Sha15,Xu18}. This is the reason they have become an essential part of the evolution towards future mobile communications. In~fact, they are one of the objects of study in the current development of fifth-generation (5G) systems~\cite{Teh14,And14,Xi15,Agi16}. 3GPP standards such as long-term evolution for machines (LTE-M), also known as enhanced MTC, and~narrow-band IoT, are only two examples of the impact that MTC are having on cellular communications~\cite{Yan17,Nok16,Els17}. Other standards proposed by different entities can be found in~\cite{Gaz17}. Coexistence with current systems will then play an important role in the entire progress of development of next mobile generations~\cite{Daw17,Ali15}.

In this framework, we can distinguish a class of MTC where a large number of devices try to access the network simultaneously, the~so-called \textit{massive} MTC (mMTC) \cite{Boc16,Eri16,Che17}. From~now on, we consider them to be sensors that collect information from the environment and send it to a central unit (CU). Unlike typical human-to-human (H2H) communications, low complexity and high energy efficiency  are preferred in mMTC instead of high data rates~\cite{Eri16}. Thus, it is crucial to look for magnitudes that measure these figures of merit reliably and, at~the same time, strategies that try to optimize them feasibly. For~instance, the authors of~\cite{Boc16} argued that non-orthogonal medium access schemes, such as sparse code multiple access, together with grant-free protocols represent a good candidate to meet the previous requirements of mMTC networks. In~this work, although~the motivation has the same origin, we focus on the communication aspects, which have an indirect impact on the complexity and energy~efficiency.

 Given that sensors transmit in a sporadic way~\cite{3GPP45820}, the~exact transmission time of each one of them may be difficult to know in some applications (for example, in~event-drive transmissions). To~facilitate the analysis, particularly in the case of a massive number of sensors, we model the sporadic transmissions of different sensors as Bernoulli random variables (RVs) with known probability. As~a result, we model the state of these devices as active or asleep (on/off).  This intermittent behavior conditions the communication between the sensors and the CU. In~particular, the~received signal at the CU from each sensor is affected by a random aggregated interference coming from the other active sensors. As~a result, a~transmission for a given sensor can be sometimes in outage, which means the interference level is high enough to make the correct detection of the signal unfeasible. In~this framework, it is then desirable to characterize the statistics of the aggregated interference to properly analyze the system~performance.

In that sense, the~outage probability, defined as the probability that the receiver is unable to decode the transmit message properly~\cite{Tse05}, represents an adequate metric of the system performance. It captures the random nature of the activity of the sensors, which is an intrinsic property of MTC networks, and~is completely defined by the statistical distribution of the aggregated interference. In addition, note that lower values of the outage probability will lead to less retransmissions and, thus, a~smaller power consumption. In~addition, the~energy used in this kind of systems during the idle state is very small compared to that when the device is active~\cite{Els17}. Hence, the~outage probability is related to the network energy efficiency and its optimization can help to improve this figure of merit. Nevertheless, as~mentioned above, an~explicit study of the energy efficiency is beyond the scope of this~paper.

On the other hand, another issue in mMTC is the medium access coordination. Given that common H2H solutions are no longer valid, a~lot of effort is put into the task of coordinating the interactions in these networks. In~particular, dedicated random access channels (RACH) are no longer feasible given the amount of signalling they need and the large number of these devices~\cite{Lie11,Gha15}. The~resulting overhead, when compared to the short packet length, yields to a reduction of the overall efficiency~\cite{Has13}. That is why strategies that control the access in a grant-free manner seem more interesting for these systems~\cite{Boc16}. However, these approaches entail a large amount of collisions and, thus, might lead to network failure. In~this work, we focus on access methods that use orthogonal resources to reduce those~collisions. 

Finally, when considering that a certain set of resources can be distributed within the network, their specific allocation is still an open, yet popular, optimization problem~\cite{Kat00}. In~this work, we~ focus on strategies that minimize the average outage probability of the sensors. Given the large number of transmitters in mMTC systems, the~solution to this problem is not~trivial.

\subsection{Prior~Work}
The statistical modeling of the aggregated interference has been studied in several cellular wireless networks considered in the literature (see, e.g., ~\cite{Rab11,Kus12,R1,R2,R3,Car10}). For~example, in~\cite{Rab11}, the authors described this distribution in the context of cognitive radio with the help of the cumulants and a truncated-stable model. Likewise, in~\cite{Kus12}, an~approximation and an analytic closed-form expression are derived for the moment generating function of the aggregated interference. Other works (e.g., \cite{R1,R2,R3}), also for cellular communications, consider the modeling of the aggregated interference in the presence of wireless channel imperfections and imperfect channel knowledge. In~these papers, the authors considered clusters and sets of base stations that cooperate in the downlink to improve the system performance and study the effect on the aggregated interference of different transmission schemes. A~more comprehensive review of modeling approaches can be found in~\cite{Car10}.

However, the~scenario under evaluation in the previous cited papers~\cite{Rab11,Kus12,R1,R2,R3,Car10} consider cellular and continuous communications, i.e.,~signals are assumed to be transmitted continuously. Therefore, the~previous works do not capture the intrinsic nature of mMTC, i.e.,~sporadic transmission and massive access, which is the core of our paper. In~fact, that is why in this work we concentrate on the randomness coming from the sensors activity to model the aggregated interference statistics. Because~of this main difference between cellular communications (continuous communications) and mMTC (sporadic transmissions), the~already existing analysis for cellular schemes cannot be applied to the mMTC scenario considered in this paper.

On the other hand, the~outage probability has been widely used to study the efficiency of sensor networks, especially using tools such as stochastic geometry~\cite{ElS13}. This field of study allows the analysis of these systems in a spatially statistical manner, i.e.,~the positions of the devices are assumed to be random following a certain distribution~\cite{Hae09,And10}. In~turn, the~activity is usually considered deterministic, which is a simplistic assumption in MTC systems. Hence, the~outage probability is formulated to capture the variations in the received signal power due only to the random positions. For~example, the authors of~\cite{Kwo13} derived analytic closed-form expressions for the signal-to-interference-and-noise (SINR), which yield to the outage probability, using homogeneous Poisson point processes for Nakagammi-$m$ and Rayleigh~fading.

To cope with the increasing number of collisions that arise with grant-free strategies, the authors of~\cite{Sen18} introduced the use of compressive sensing in the context of mMTC systems. In~\cite{Kel17}, the authors proposed a scheme based on the distribution of grants through a multi-antenna technology. An~algorithm relying on the maximization of the random access efficiency and the estimation of the number of devices is described in~\cite{Oh15}. In~\cite{Si15}, the authors presented a methodology relying on queues and the observed traffic load to guarantee a certain statistical quality of service (QoS). An~overview of more alternatives can be found in~\cite{Lay14}. 

Regarding the resource allocation problem, different methodologies can be found within the literature. The authors of \cite{Lag16} proposed a graph-based method to optimize the maximum average resource utilization in the network. Differently, a~dynamic scheduling solution that relies on devices priorities and that takes into account the mMTC scenario is presented in~\cite{Sal18}, where the impact on the outage probability is also analyzed. Moreover, in~\cite{Alc18}, the authors analyzed the scheduling of resources with the help of non-orthogonal multiple access using stochastic geometry. A~similar point of view can be found in~\cite{Guo17}, where random and channel-aware allocation strategies are also investigated. For~more approaches, refer to the survey in~\cite{Xia18} (and references~therein).

\subsection{Contributions}
With the above considerations, this paper can be seen as an extension of the work presented by the same authors in the conference paper~\cite{SLMGS18}. In~particular, the~main contributions of this work are the~following:
\begin{itemize}
    \item[(1)]{Several analytic closed-form approximations of the statistics of the aggregated interference that include the random activity of the sensors. We present their advantages over the ones introduced in~\cite{SLMGS18} by the same authors.}
\end{itemize}

Our approach is firstly formulated for a generic scenario and is later particularized to an uplink multi-antenna mMTC setup in order to present the second novelty of this~work:
\begin{itemize}
    \item[(2)]{An analytic closed-form expression for the outage probability of the sensors. It is used to evaluate the performance of this type of systems.}
\end{itemize}

Finally, in~this work, we also address the following~issues:
\begin{itemize}
    \item[(3)]{A medium access scheme with spatial beamforming.}
    \item[(4)]{A graph-based resource allocation strategy that minimizes the average outage probability.}
\end{itemize}
 
\subsection{Organization} 
The paper is organized as follows. In~Section~\ref{sec:2}, the~system model is described. Next, the~approximations of the aggregated interference statistics are presented in Section~\ref{sec:3}. The~outage probability is derived in Section~\ref{sec:4} for an illustrative reference scenario. Next, the~resource allocation problem is formulated and solved in Section~\ref{sec:5}. Numerical results are shown in Section~\ref{sec:6} to illustrate the accuracy of the approximations and evaluate the performance of the proposed resource allocation strategy. Section~\ref{sec:7} is devoted to~conclusions.

\subsection{Notation} 
In this paper, scalars are denoted by italic letters. Boldface lowercase and uppercase letters denote vectors and matrices, respectively. For~a given vector or matrix, the~operations $(\cdot)^{\textrm{T}}$ and $(\cdot)^{\textrm{H}}$ denote their transpose and Hermitian, respectively. Matrix $\mathbf{I}_M$ denotes the identity matrix of size $M \times M$. For~given sets $\mathcal{A}$ and $\mathcal{B}$, the~union and intersection are denoted by $\mathcal{A} \cup \mathcal{B}$ and $\mathcal{A} \cap \mathcal{B}$, respectively. The~cardinality of $\mathcal{A}$ is denoted by $\vert \mathcal{A} \vert$. $\mathbb{C}^{m \times n}$ and $\mathbb{N}^{m \times n}$ denote the $m$ by $n$ dimensional complex space and natural space, respectively. The~Gaussian and the Gaussian circularly symmetric complex distributions are denoted by $\mathcal{N}(\cdot,\cdot)$ and $\mathcal{N}_{\mathbb{C}}(\cdot,\cdot)$, respectively. The Bernoulli distribution with parameter $p$ is denoted by \textit{Ber}($p$). Besides, for~the sake of clarity in the explanation, in~Table~\ref{tab:1}, we~include a list with most of the relevant variables used in further sections of this~work.

\section{System~Model} \label{sec:2}
In this paper, we start by considering a generic scenario in which a set of sensors transmit towards a CU through possibly different orthogonal resources. Then, only the sensors sharing the same resources will interfere each other. It is important to highlight that, under~these first few assumptions, the~upcoming analysis applies to many network~topologies.

To model the activity of the sensors, we consider them to be in active or sleep mode (on/off) with the help of a Bernoulli RV $\beta_j \sim$ \textit{Ber}($p_j$), where $j$ refers to the sensor index, and~$p_j$ represents the probability that sensor $j$ is active and transmitting. Besides, we assume that the RVs corresponding to different sensors are independent. Note that this practice is commonly used in the literature to describe the sporadic nature of transmissions (cf.~\cite{Rab11,Kus12,Car10}).

At the detection stage, the~received signal at the CU from each sensor is affected by an aggregated interference coming from the other active sensors, which follows a certain statistical distribution. The~corresponding probability mass function (pmf) of the aggregated interference is the result of the sum of scaled independent Bernoulli~RVs.

Let us consider $i$ as the index of the sensor under analysis, the~transmit signal of which is to be detected. Besides, to~represent actual communication from the $i$th sensor perspective, we consider that this sensor is active. Hence, its activity variable $\beta_i$ is set to one, i.e.,~$\beta_i = 1$. 

With the above considerations, the~SINR corresponding to the received signal at the CU from sensor $i$ when it is transmitting~is 
\begin{equation}
    \rho_i = \frac{a_{i,i}}{\sigma_{n,i}^2 + \Gamma_i},
\label{eq:1}
\end{equation}

\begin{table}[H]
\caption{Summary of Variable~Notation.}
\begin{center}
    \begin{tabular}{|c|c|}
        \hline
        \textbf{Notation} & \textbf{Description} \\
        \hline
        $\beta_j$ & Bernoulli random variable accounting for the activity of sensor $j$ \\ 
        \hline
        $p_j$ & Probability that sensor $j$ is active and transmitting\\ \hline
        $\rho_i$ & SINR of the received signal from sensor $i$ \\ \hline 
        $a_{i,i}$ & Received power of the signal from sensor $i$ \\ \hline
        $a_{j,i}$ & Received power of the signal from sensor $j$ at the detector of sensor $i$ \\ \hline 
        $\sigma_{n,i}^2$ & Power of the noise at the detector of the signal from sensor $i$ \\ \hline
        $\mathcal{J}_i$ & Set of sensors interfering to sensor $i$ \\ \hline 
        $\Gamma_i$ & Aggregated interference that the signal from sensor $i$ perceives \\ \hline
        $P_{\textrm{out}}^i$ & Outage probability of sensor $i$ \\ \hline 
        $p_{\Gamma_i}(\gamma_i)$ & Probability mass function of the aggregated interference \\ \hline
        $\varphi_{\Gamma_i}(t)$ & Characteristic function of the aggregated interference \\ \hline 
        $\phi(\gamma_i;\mu_i,\sigma_i)$ & \shortstack{Probability density function of the Gaussian distribution \\ with mean $\mu_i$ and standard deviation $\sigma_i$} \\ \hline
        $\mu_i$ & Mean of the aggregated interference \\ \hline 
        $\sigma_i$ & Standard deviation of the aggregated interference \\ \hline
        $\kappa_n^i$ & $n$th cumulant of the aggregated interference \\ \hline 
        $B_n$ & $n$th Bell polynomial \\ \hline
        $H_n$ & $n$th Hermite polynomial \\ \hline 
        $\mu_{i,n}'$ & $n$th non-centralized order moment of the aggregated interference \\ \hline
        $f_X(x;\mu,\sigma,a,b)$ & \shortstack{Probability density function of the truncated Gaussian distribution \\ with mean 
        $\mu$ and standard deviation $\sigma$, defined between $a$ and $b$}
        \\ \hline 
        $\phi_s(\cdot)$ & Probability density function of the standard Gaussian random variable \\ \hline
        $\Phi_s(\cdot)$ & Cumulative distribution function of the standard Gaussian random variable \\ \hline 
        $\delta$ & Detection threshold \\ \hline
        $M$ & Number of sensors \\ \hline 
        $P$ & Transmit power of the sensors \\ \hline
        $N$ & Number of receive antennas \\ \hline 
        $L$ & Number of spatial beams \\ \hline
        $K$ & Number of collector nodes \\ \hline 
        $R$ & Number of available orthogonal resources \\ \hline
        $\mathcal{J}_{k,l}$ & Sensors detected at the beam $l$ of the collector node $k$ \\ \hline 
        $\mathcal{T}_{k,l}$ & Resources allocated at the beam $l$ of the collector node $k$ \\ \hline
        $\Gamma_i^{\textrm{intra}}$ & Intra-beam aggregated interference \\ \hline 
        $\Gamma_i^{\textrm{inter}}$ & Inter-beam aggregated interference \\ \hline
        $\mathcal{J}_i^{\textrm{intra}}$ & Set of sensors interfering to sensor $i$ and detected at the same beam \\ \hline 
        $\mathcal{J}_i^{\textrm{inter}}$ & Set of sensors interfering to sensor $i$ and detected at another beam \\ \hline
        $R_{\textrm{dep}}$ & Radius determining the deployment of sensors \\ \hline 
        $R_{\textrm{int}}$ & Radius determining the range of the interference \\ \hline
    \end{tabular} 
\end{center}
\label{tab:1}
\end{table}

\noindent
where $a_{i,i}$ is used to denote the received power of the signal from sensor $i$, i.e.,~the gain of the channel multiplied by the transmit power. The~second sub-index indicates the detector in charge of detecting the signal from sensor $i$. In~fact, $\sigma_{n,i}^2$ corresponds to the power of the noise at the detector where the signal from sensor $i$ is detected. Finally, the~term $\Gamma_i$ represents the aggregated interference:
\begin{equation}
    \Gamma_i = \sum_{j \in \mathcal{J}_i} \beta_j a_{j,i},
\label{eq:2}
\end{equation}
where $\mathcal{J}_i$ is the set of sensors that can potentially interfere with sensor $i$, i.e.,~those that use the same resources as the ones used by sensor $i$. Likewise, $a_{j,i}$ represents the received power of the signal from sensor $j$ at the detector where the signal coming from sensor $i$ is~detected.

The distribution of the aggregated interference $\Gamma_i$ can be obtained through a set of discrete convolutions given that all the individual addends $\beta_j a_{j,i}$ are binary and independent~\cite{SLMGS18}. However, such operation can be tedious since the complexity depends on the number of interfering devices $\vert \mathcal{J}_i \vert$ and grows exponentially with this magnitude. Thus, it becomes rapidly unfeasible, even for a small $\vert \mathcal{J}_i \vert$. Even if a Monte-Carlo based numerical approach were always available, it would still carry a large computational complexity. This is why, in the following, we propose two different and reasonable alternatives to express the previous statistics using closed-form approximations. In~addition, thanks to having these closed-form approximations, and~not only numerical ones, an~optimization of the resource allocation can be carried out, as~shown~ below.

\section{Approximations of the Statistics of the Aggregated~Interference} \label{sec:3}
The purpose of this work is to provide a closed-form expression approximating the pmf of the aggregated interference $\Gamma_i$, i.e.,~$p_{\Gamma_i} (\gamma_i)$. To~this end, we propose two alternatives. The~first one is based on the characteristic function of $\Gamma_i$, whereas the second one relies on the Gram--Charlier series expansion of a Gaussian kernel. The~latter is used in Section~\ref{sec:4} to derive an analytic closed-form expression for the outage probability (i.e., the probability of not being able to decode the sensors transmit signal because the corresponding SINR at the detection stage is lower than a certain predefined threshold). This is the main novelty of the paper and represents the core of our work. To~the best of our knowledge, no similar work has been done in this~direction.

The approximations for the statistics of the aggregated interference that we develop in this section could be used in many different applications. In~the following, we list some illustrative~examples:
\begin{itemize}
    \item {\textit{Outage Probability}}: Thanks to the approximation of the interference statistics, we are able to find an analytic closed-form expression for the outage probability (defined as the probability that the SINR is below a certain detection threshold). Based on that, and~considering an orthogonal multiple access with limited number of available resources, an~allocation scheme could be designed to minimize this metric improving, therefore, the~system performance.
    \item {\textit{Throughput}}: The approximated statistics of the aggregated interference could also be useful to obtain an analytic closed-form expression for the throughput of the sensors (e.g., through Shannon's capacity). Note that this is indeed related to the outage probability, yet it captures different aspects of the communication. 
    \item {\textit{Harvested Energy}}: Given the large number of active devices in mMTC, the~energy coming from the transmitted signals can be recycled (i.e., harvested). Then, we could employ the derived statistics to characterize the amount of harvested energy.
    \item {\textit{Power Consumption}}: Taking into account the number of retransmissions (e.g., due to a high outage probability) and the different energy supplies (e.g., harvested energy), a~power consumption model of the sensors could be derived to study and optimize the management of available energy within the network.
\end{itemize}

As already mentioned, these are only few examples of possible uses of the approximations of the interferences statistics that are developed in this work. In~particular, in~this paper, we concentrate on the first application, which is first discussed in Section~\ref{sec:4} (outage probability) and then in Section~\ref{sec:5} (resource allocation). The~rest might be object of study of future works.

\subsection{Characteristic~Function} \label{sec:3.1}
To obtain the pmf of $\Gamma_i$, we can use the characteristic function. For~$X \sim$ \textit{Ber}($p$), its characteristic function reads~as
\begin{equation}
    \varphi_X (t) = \textrm{E}[e^{\mathsf{i}tX}] = 1 - p + pe^{\mathsf{i}t},
\label{eq:3}
\end{equation}
where $\mathsf{i}$ denotes the square root of $-1$ (not to be confused with the sensor index $i$). Then, when introducing the weights $a_{j,i}$ with the RVs $\beta_j$, the~total characteristic function of $\Gamma_i$ results
\begin{equation}
    \varphi_{\Gamma_i} (t) = \prod_{j \in \mathcal{J}_i} (1 - p_j + p_j e^{\mathsf{i}ta_{j,i}}),
\label{eq:4}
\end{equation}
assuming independence among individuals. Since this can be interpreted as the Fourier transform of the pmf (with opposite sign), we can just invert this transformation to obtain the pmf:
\begin{equation}
    p_{\Gamma_i}(\gamma_i) = \mathcal{F}^{-1}\{\varphi_{\Gamma_i} (t)\} = \frac{1}{2 \pi}\int_{2 \pi} \varphi_{\Gamma_i} (t) e^{-\mathsf{i}t \gamma_i} dt.
\label{eq:5}
\end{equation}

This way, we go from a set of convolutions to simple products and the inverse Fourier transform, which can be calculated numerically with the inverse fast Fourier transform (IFFT). The~number of operations are then significantly reduced and, given that the number of points used in the IFFT for the discretization of the continuous inverse Fourier transform is actually limited, an approximation 
of the pmf can now be obtained. In fact, note that the necessary points for good precision in the IFFT increase with the number of interfering devices $\vert \mathcal{J}_i \vert$. This can be translated into a high computational cost, yet bearable in finite time. As~a result, assuming that enough points are used in the IFFT routine, this method is only used to validate the accuracy of the next alternative. Finally, we can relate this approach to the moment generating function presented in~\cite{Kus12} in the sense that it describes a similar transformation of the statistical distribution of the aggregated interference, i.e.,~it can be interpreted as the Laplace transform of the pmf.

\subsection{Gram--Charlier~Approximation} \label{sec:3.2}
Another way to find a suitable and more computationally efficient expression for the pmf of the aggregated interference $\Gamma_i$ is by means of a continuous approximation. To~this end, in~this work, we make use of the Gram--Charlier series expansion of a Gaussian kernel~\cite{Bre17}:
\begin{equation}
    p_{\Gamma_i}(\gamma_i) = \phi(\gamma_i;\mu_i,\sigma_i) \sum_{n = 0}^{\infty} \frac{1}{n ! \sigma_i^n} B_n(0,0,\kappa_3^i, \ldots, \kappa_n^i) H_n (\bar{\gamma}_i),
\label{eq:6}
\end{equation}
where $\phi(\gamma_i;\mu_i,\sigma_i)$ is the probability density function (pdf) of the Gaussian distribution with mean $\mu_i$ and standard deviation $\sigma_i$, and~$\kappa_n^i$, $B_n$, and $H_n$ are the $n$th cumulant of $\Gamma_i$, Bell, and Hermite polynomials, respectively~\cite{Com12}. The~term $\bar{\gamma}_i = (\gamma_i - \mu_i)/\sigma_i$ is the normalized~argument.

This approach allows the approximation of a distribution through its statistical moments. Given that these magnitudes can be determined for $\Gamma_i$, this inference method represents a good option for approximating its pmf. Since all $\beta_j$ are independent, the~first two moments of $\Gamma_i$ are~indeed
\begin{equation}
    \mu_i = \sum_{j \in \mathcal{J}_i } p_j a_{j,i}, \quad 
    \sigma_i^2 = \sum_{j \in \mathcal{J}_i } p_j (1 - p_j) a_{j,i}^2.
\label{eq:7}
\end{equation}

The expression in~Equation \eqref{eq:6} leads to a perfect approximation for infinite addends as it converges to the actual distribution~\cite{Bre17}. Since adding more terms reduces the error in the approximation, we truncate the series up to a finite number of addends to work with a tractable expression. Thereby, similar to~the work described in~\cite{Rab11}, we obtain an expression based on the cumulants of the aggregated interference, which can be found recursively in terms of the first $n$th non-centralized order moments $\mu_{i,n}' = \sum_{j \in \mathcal{J}_i} p_j a_{j,i}^n$ \cite{Smi95}:
\begin{equation}
    \kappa_n^i = \mu_{i,n}' - \sum_{m = 1}^{n - 1}\binom{n - 1}{m - 1} \kappa_m^i \mu_{i,n-m}'.
\label{eq:8}
\end{equation} 

It is important to highlight that other kernels can be employed for the expansion. The~reason for choosing the Gaussian kernel follows from the reasoning in~\cite{SLMGS18}, where the authors justified, through the proof of Lyapunov's central limit theorem (CLT), the~use of a Gaussian distribution to properly approximate the interference statistics. In~fact, note that~\cite{SLMGS18} presented a particular case of the Gram--Charlier series expansion defined in~Equation \eqref{eq:6} when the order is set to $0$. However, the~approach in~\cite{SLMGS18} (i.e., order $0$) might not be adequate for some scenarios as the necessary condition might not be fulfilled (see~Expression (15) in~\cite{SLMGS18}). This is the case when the number of interfering devices is not sufficiently large or when sensors have different transmission probabilities, in~which case Expression (15) in~\cite{SLMGS18} yields
\begin{equation}
    \frac{1}{\sigma_i^{2 + \epsilon}} \sum_{j \in \mathcal{J}_i} a_{j,i}^{2 + \epsilon} p_j (1 - p_j).
\label{eq:9}
\end{equation}

According to~\cite{SLMGS18}, Equation \eqref{eq:9} tends to zero for equal $p_j$ not close to zero and a large number of interfering devices $\vert\mathcal{J}_i\vert$. In~that case, Condition (15) in~\cite{SLMGS18} is satisfied and the Gaussian kernel with order set to $0$ suffices. However, since here the probabilities are different and can be small, the~simple Gaussian distribution may not be enough for good accuracy. That is why we propose an extension of that approach, i.e.,~an expansion of a Gaussian kernel with order higher than $0$.

Until now, we have assumed that the pmf of $\Gamma_i$ has an infinite support. However, it is actually lower and upper bounded by $0$ and $J_i = \sum_{j \in \mathcal{J}_i} a_{j,i}$, respectively. These are the extreme possible values for the aggregated interference. For~analytic consistency, we must use a truncated Gaussian kernel~\cite{Bur14}. Let $X \sim \mathcal{N}(\mu,\sigma^2)$ be a Gaussian RV defined between $a$ and $b$ with~pdf
\begin{equation}
    f_X(x; \mu, \sigma, a,b) = \frac{\phi_s(\frac{x - \mu}{\sigma})}{\sigma \big(\Phi_s(\frac{b - \mu}{\sigma})- \Phi_s(\frac{a - \mu}{\sigma}) \big)} \\
    = \frac{1}{\sigma F}\phi_s\Big(\frac{x - \mu}{\sigma}\Big) = \frac{1}{F} \phi(x; \mu,\sigma),
\label{eq:10}
\end{equation}
in the interval $a \leq x \leq b$ and 0 otherwise. The~term $F$ represents the normalization factor used to achieve unit area, i.e.,~$F = \Phi_s(\frac{b - \mu}{\sigma})- \Phi_s(\frac{a - \mu}{\sigma})$. Note that $\phi_s(\cdot) \equiv \phi(\cdot,0,1)$ refers to the pdf of the standard Gaussian RV and $\Phi_s(\cdot)$ is the corresponding cumulative distribution function (cdf).

In our case, by~defining $F_i = \Phi_s(\frac{J_i - \mu_i}{\sigma_i})- \Phi_s(\frac{0 - \mu_i}{\sigma_i})$, we have that the new kernel reads~as
\begin{equation}
    f_{\Gamma_i}(\gamma_i; \mu_i, \sigma_i, 0,J_i) = \phi(\gamma_i;\mu_i,\sigma_i)/F_i,
\label{eq:11}
\end{equation}
for $0 \leq \gamma_i \leq J_i$. Thereby, it can be shown that a similar expansion can be found for a finite support. We~only need to introduce the cumulants of the truncated Gaussian distribution since, differently from before, they are not zero for orders higher than $2$ and the first two do not equal $\kappa_1^i$ and $\kappa_2^i$, i.e.,
\begin{equation}
p_{\Gamma_i}(\gamma_i) = \frac{1}{F_i} \phi(\gamma_i;\mu_i,\sigma_i) \sum_{n = 0}^{\infty} \frac{1}{n ! \sigma_i^n} B_n(\eta_1^i,\ldots, \eta_n^i) H_n (\bar{\gamma}_i),
\label{eq:12}
\end{equation}
where we define $\eta_n^i = \kappa_n^i- \lambda_n^i$ with $\lambda_n^i$ being the cumulants of the new truncated kernel. As~ mentioned, they are related to the non-centralized moments, which can be obtained in an analytic closed-form manner~\cite{Bur14}. In~particular, for~a general $X \sim \mathcal{N}(\mu,\sigma^2)$ with support $[a,b]$, we~have
\begin{equation}
    \mu_{k}' = \sum_{i = 0}^k \binom{k}{i}\sigma^i\mu^{k - i} L_i,
\label{eq:13}
\end{equation}
where $L_i = (\bar{a}^{i-1}\phi_s(\bar{a}) - \bar{b}^{i-1}\phi_s(\bar{b}))/F + (i-1)L_{i-2}$, $\bar{a} = (a - \mu)/\sigma$, $\bar{b} = (b - \mu)/\sigma$ and $L_0 = 1$.

Finally, when truncating the series up to order $N$, we obtain the continuous approximation for the pmf of the aggregated interference, denoted by $f_{\Gamma_i}^N(\gamma_i)$. For~instance, for~$N = 5$, we have:
\begin{equation}
    p_{\Gamma_i}(\gamma_i) \approx \nu_i (\bar{\gamma}_i) \phi(\gamma_i;\mu_i,\sigma_i)/ F_i = f_{\Gamma_i}^5(\gamma_i),
\label{eq:14}
\end{equation}
where
\begin{equation}
    \nu_i(\bar{\gamma}_i) = 1 + \frac{\eta_1^i}{\sigma_i}H_1(\bar{\gamma}_i) + \frac{\eta_2^i}{2\sigma_i^2} H_2(\bar{\gamma}_i) + \frac{\eta_3^i}{6\sigma_i^3} H_3 (\bar{\gamma}_i) + \frac{\eta_4^i}{24 \sigma_i^4} H_4 (\bar{\gamma}_i) + \frac{\eta_5^i}{120\sigma_i^5} H_5 (\bar{\gamma}_i).
\label{eq:15}
\end{equation}

In the next section, this approximation is used to express the outage probability of the sensors (i.e., the probability of the SINR being below a certain predefined threshold). Then, in~Section~\ref{sec:5}, this expression is used to design a resource allocation~strategy.

\section{Outage~Probability} \label{sec:4}
This section is devoted to presenting an analytic closed-form expression for the outage probability taking into account the massive uplink (UL) communication in mMTC systems. This figure of merit represents the probability that the receiver is unable to decode the transmit message and, thus, it can also be interpreted as the portion of time for which the communication fails. Thereby, this magnitude can be used as a valuable performance indicator in this kind of systems (cf.~\cite{Car10}). Furthermore, recall that lower values of the outage probability result in less retransmissions and, thus, in~a lower power consumption and also in lower delays. That is why, in Section~\ref{sec:5}, for~a better efficiency, we aim to find an allocation that minimizes this~metric.

The outage probability is defined as~\cite{Kwo13}:
\begin{equation}
    P_{\textrm{out}}^i = \textrm{Pr}\{\rho_i < \delta\},
\label{eq:16}
\end{equation}
where $\rho_i$ is the SINR of the signal from sensor $i$ defined in~Equation \eqref{eq:1}, and~$\delta$ is the detection threshold, i.e.,~the~minimum required SINR for the message to be successfully decoded without retransmission. Hence, we have an outage whenever the quality of the communication is not enough for correct decoding. This may happen in cases of large interference, either due to high channel gains and/or high activities of non-intended sources (represented by the set of interfering sensors $\mathcal{J}_i$). 

As mentioned above, we use the Gram--Charlier approximation derived in Section~\ref{sec:3.2} to find an analytic closed-form expression for the outage probability defined in Equation \eqref{eq:16}. Note that this magnitude may be reduced when mitigating the interference. Therefore, we consider a scenario with a set of multi-antenna collector nodes (CNs), each equipped with a set of predefined spatial beams. This way, we can help the CU overcome the problem of massive access and obtain a reasonable outage probability $P_{\textrm{out}}^i$.

For the sake of clarity in the explanation, we start our analysis from a simple setup, and~we then sophisticate it towards the more generalized multiple CN multi-antenna scenario. Note that each scenario is a particular case of the following one, but~we have decided to present the setups in this constructive way to avoid any possible confusion in the description of the notation and the developments. Thus, all previous setups are particular cases of the last scenario. In~that sense, although~at each step we characterize the system model, the~expression of $P_{\textrm{out}}^i$ is only presented for the general case in order to avoid~redundancy.

\subsection{Single-Antenna~CU} \label{sec:4.1}
Let us consider a single-antenna CU collecting the information from a set of $M$ transmitting single-antenna sensors using the same access resources. In~that case, the~received signal~is
\begin{equation}
    y = \sum_{j = 1}^M h_j \beta_j x_j + w \in \mathbb{C},
\label{eq:17}
\end{equation}
where $h_j \in \mathbb{C}$ is the channel of sensor $j$, $x_j$ is the transmit signal with zero mean and power $P$, independent for each sensor, and~$w \sim \mathcal{N}_{\mathbb{C}}(0,\sigma_{w}^2)$ is the additive noise, independent of $x_j$.

Thereby, since no further processing is performed at the CU, the~received power $a_{i,i}$ of the signal from sensor $i$ defined in Section~\ref{sec:2} and~the addends $a_{j,i}$ in the aggregated interference $\Gamma_i$ from~Equation~\eqref{eq:2} read~as
\begin{equation}
    a_{i,i} = P \vert h_i \vert^2, \quad a_{j,i} = P \vert h_j \vert^2,
\label{eq:18}
\end{equation}
and, given that all sensors share the same resources, the~interfering set results $\mathcal{J}_i = \{j \neq i\}$. In~addition, in~this case, the noise at the detection stage is the same for all sensors, i.e.,~$\sigma_{n,i}^2 = \sigma_{w}^2$.

\subsection{Multiple-Antenna~CU} \label{sec:4.2}
To reduce the interference coming from the rest of sensors, we now consider that the CU is equipped with $N$ antennas. The~received signal is given~by
\begin{equation}
    \bm{r} = \bm{H} \bm{\beta} \bm{x} + \bm{w} \in \mathbb{C}^{N},
\label{eq:19}
\end{equation}
where $\bm{w} \sim \mathcal{N}_{\mathbb{C}}(\mathbf{0},\sigma_w^2 \mathbf{I}_N)$ is the corresponding noise vector and $\bm{x} = [x_1,\ldots,x_M]^{\textrm{T}} \in \mathbb{C}^M$ is the vector containing all the different transmit signals $x_j$ with zero mean and power $P$, independent of the noise $\bm{w}$ and independent among them. In~addition, $\bm{H} = [\bm{h}_{1},\ldots, \bm{h}_{M}] \in \mathbb{C}^{N \times M}$ is the matrix containing the set of individual channels $\bm{h}_j$ of each sensor with respect to (w.r.t.) the CU, and~$\bm{\beta} = \textrm{diag}(\beta_1,\ldots,\beta_M)$ is the matrix containing the different RVs $\beta_j$ of all~sensors.

Given the degrees of freedom provided by the multi-antenna technology, linear processing is employed in this setup. More specifically, the CU now has $L$ predefined spatial beams implemented through a spatial filter represented by the matrix $\bm{G} = [\bm{g}_{1},\ldots,\bm{g}_{L}] \in \mathbb{C}^{N \times L}$. A possible design option is to construct the different beams so that their pointing directions are equispaced. However, for~the sake of generality, we keep the filtering scheme generic and represented by $\bm{G}$.

The signal coming from each sensor is then detected at the output of a given spatial beam. The~$L$ outputs at the $L$ beams can be collected in a single vector given by:
\begin{equation}
    \bm{y} = \bm{G}^{\textrm{H}} \bm{r} = \bm{G}^{\textrm{H}} \bm{H} \bm{\beta} \bm{x} + \bm{n},
\label{eq:20}
\end{equation} 
where $\bm{y} = [y_{1},\ldots,y_{L}]^{\textrm{T}} \in \mathbb{C}^{L}$ is the processed signal and $\bm{n} = \bm{G}^{\textrm{H}} \bm{w} \in \mathbb{C}^{L}$ is the filtered~noise. 

In this work, and~for the sake of simplicity, we assume that the signal from a given sensor is decoded using the output signal of a single beam. Then, to~determine which is the detecting beam for each sensor, a~possible criterion is to choose the one leading to the largest signal-to-noise ratio (SNR) at the output of the spatial beam. Intuitively, it represents the beam where the quality of the signal might be better. Let $l(i)$ represent the index of the beam used to detect the signal coming from the $i$th sensor:
\begin{equation}
    l(i) = \underset{l }{\textrm{argmax}} \, \frac{\vert \bm{g}_{l}^{\textrm{H}} \bm{h}_{i} \vert^2}{\sigma_w^2 \| \bm{g}_{l} \|_2^2},
\label{eq:21}
\end{equation}
where $\| \bm{g}_{l} \|_2$ denotes the L2-norm of the filter $\bm{g}_{l}$. Therefore, we denote $\bm{g}_{l(i)}$ as the spatial filter used for the detection of sensor $i$. Note that, if a different beam selection strategy is employed, we have a different expression for $l(i)$. For~the sake of generality, we just use $l(i)$; hence, the~upcoming analysis holds regardless of the beam selection~criterion.

Accordingly, we can express the term $a_{i,i}$ for the received power of sensor $i$ and the interfering power terms $a_{j,i}$ as
\begin{equation}
    a_{i,i} = P \vert \bm{g}_{l(i)}^{\textrm{H}} \bm{h}_{i} \vert^2, \quad
    a_{j,i} = P \vert \bm{g}_{l(i)}^{\textrm{H}} \bm{h}_{j} \vert^2,
\label{eq:22}
\end{equation} 
and, assuming that there is still a complete reuse of resources, i.e.,~all sensors share the same ones, the~interfering set is again $\mathcal{J}_i = \{j \neq i\}$. In~addition, the~power of the noise affecting the $i$th sensor is given by $\sigma_{n,i}^2 = \sigma_w^2 \| \bm{g}_{l(i)} \|_2^2$. Finally, note that the second sub-index in $a_{i,i}$, $a_{j,i}$ and $\sigma_{n,i}^2$ is used to indicate where the $i$th sensor is actually detected, i.e.,~at the beam $l(i)$.

\subsection{Multiple-Antenna CNs and~CU} \label{sec:4.3}
\begin{figure}[t]
\centerline{\includegraphics[scale = 0.75]{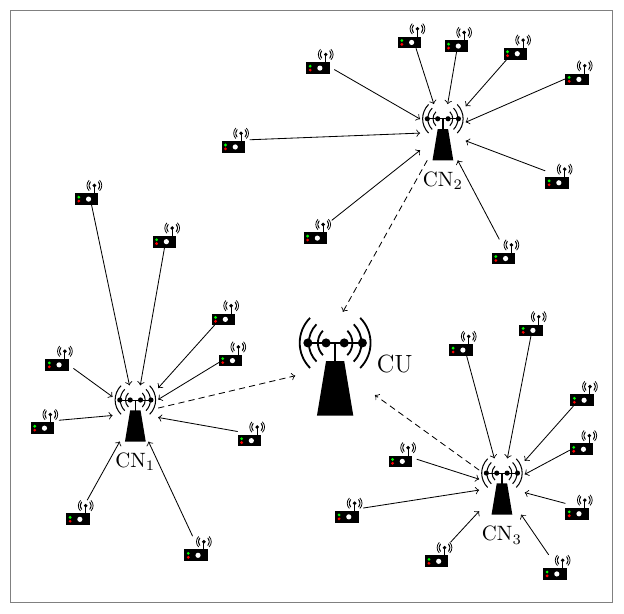}}
\caption{System setup for $K = 3$, $N = 4$ and $M_k = 9 \, \forall k$.}
\label{fig:1}
\end{figure}

To further enhance the performance of the system, we now consider an extension of the previous setup with $K$ data CNs, each one also equipped with $N$ antennas. Each of them is responsible for collecting the information from a subset of $M_k$ single-antenna sensors for later retransmitting it to a CU, with~$M = \sum_{k = 1}^K M_k$. Thereby, we have a two-hop communication system, as~illustrated in Figure~\ref{fig:1}. Such multi-hop scheme is largely exploited in the literature~\cite{Guo17,Sal18} and in standards such as LTE-M~\cite{Zhe12,Gha15}. In~this work, we focus on the communication in the first hop (solid line) and leave the second stage (dashed line) for further~studies.

Each CN has $L$ predefined spatial beams, represented with the spatial filter matrix $\bm{G}_k = [\bm{g}_{k,1},\ldots,\bm{g}_{k,L}] \in \mathbb{C}^{N \times L}$, where $k$ represents the CN index, i.e.,~$k = \{1,\ldots,K\}$. 

Now, the~signal coming from each sensor is detected at the output of a given spatial beam of a given CN, i.e.,~at a given CN--beam pair or tuple. Besides, we assume that there is no cooperation among beams and no cooperation among CNs in the signal detection~stage. 

The signal received at the $k$th CN can be written~as
\begin{equation}
    \bm{r}_k = \bm{H}_k \bm{\beta} \bm{x} + \bm{w}_k \in \mathbb{C}^N,
\label{eq:23}
\end{equation}
where $\bm{w}_k \sim \mathcal{N}_{\mathbb{C}}(\mathbf{0},\sigma_w^2\mathbf{I}_N)$ is the noise vector and $\bm{H}_k = [\bm{h}_{k,1},\ldots, \bm{h}_{k,M}] \in \mathbb{C}^{N \times M}$ is the matrix containing the channels of each sensor w.r.t. each CN, i.e.,~$\bm{h}_{k,i}$ is the channel between sensor $i$ and the $k$th CN. Again, $\bm{x} \in \mathbb{C}^M$ is the vector of independent transmit signals with zero mean and power $P$, independent of $\bm{w}_k$, and~$\bm{\beta} = \textrm{diag}(\beta_1,\ldots,\beta_M)$ is the matrix of RVs $\beta_j$.

The $L$ outputs at the $L$ beams of the $k$th CN can be collected in a single vector given by:
\begin{equation}
    \bm{y}_k = \bm{G}_k^{\textrm{H}} \bm{r}_k = \bm{G}_k^{\textrm{H}} \bm{H}_k \bm{\beta} \bm{x} + \bm{n}_k,
\label{eq:24}
\end{equation} 
with $\bm{y}_k = [y_{k,1},\ldots,y_{k,L}]^{\textrm{T}} \in \mathbb{C}^{L}$ the processed signal at the $k$th CN and $\bm{n}_k = \bm{G}_k^{\textrm{H}} \bm{w}_k \in \mathbb{C}^{L}$ the filtered noise. The~channels and filters are specified in Section~\ref{sec:6} for an exemplifying~scenario. 

Similar to above, for~each sensor, we focus on the detection at the CN--beam pair leading to the largest SNR after the spatial filter. Recall that other criteria to choose the CN--beam pair could also be used. Let now $k(i)$ and $l(i)$ represent the indexes of the CN and the beam used to detect the signal coming from the $i$th sensor using the previous criterion:
\begin{equation}
    (k(i),l(i)) = \underset{(k,l) }{\textrm{argmax}} \, \frac{\vert \bm{g}_{k,l}^{\textrm{H}} \bm{h}_{k,i} \vert^2}{\sigma_w^2 \| \bm{g}_{k,l} \|_2^2},
\label{eq:25}
\end{equation}
where $\| \bm{g}_{k,l} \|_2$ denotes the L2-norm of the filter $\bm{g}_{k,l}$. Hence, $\bm{g}_{k(i),l(i)}$ represents the spatial filter used for the detection of sensor $i$. In~addition, the~noise power is given by $\sigma_{n,i}^2 = \sigma_w^2 \| \bm{g}_{k(i),l(i)} \|_2^2$.

On the other hand, note that, until now, we have considered all sensors to interfere among them. In~other words, we have assumed that they are all using the same orthogonal resource. Nevertheless, in~real systems, more than one resource is usually available. Therefore, from~now on, we consider that we dispose of $R$ orthogonal resources with indexes $\{1,\ldots,R\}$, and~that they are allocated to the different CN--beam~tuples.

In that sense, the~sensor that is detected at a certain CN--beam pair uses the resources allocated to that tuple. For~the moment, we consider that sensors know which resources they can employ, and~we ignore the way this information is acquired. An~example of a mechanism that provides this knowledge at the sensors side is described in Section~\ref{sec:6}.

Furthermore, we distinguish between the case where only one resource is allocated to each CN--beam pair and the situation where these tuples can use more than one resource. They are referred to as \textit{single}- and \textit{multiple-resource} scenario, respectively. In~the latter, each sensor chooses one of the resources available for their pair, defined in~Equation \eqref{eq:25}, at~random. In~both setups, resources are allocated to reduce the impact of the interference, as~discussed in Section~\ref{sec:5}.

With the above considerations, and~considering $\beta_i = 1$, the~received signal from the $i$th sensor at the tuple $(k(i),l(i))$ can be written as (cf.~Equation \eqref{eq:24})
\begin{equation}
    y^{(i)} = \bm{g}_{k(i),l(i)}^{\textrm{H}} \bm{h}_{k(i),i} x_i + I_i + n^{(i)},
\label{eq:26}
\end{equation}
where $y^{(i)} \equiv y_{k(i),l(i)} \in \mathbb{C}$ is the received signal; $\bm{g}_{k(i),l(i)}$ and $\bm{h}_{k(i),i}$ are the filter and channel of sensor $i$, respectively; and~$n^{(i)} \equiv \bm{g}_{k(i),l(i)}^\textrm{H}\bm{n}_{k(i)} \in \mathbb{C}$ is the noise at that tuple, with~power $\sigma_{n,i}^2$.

Each of the $y^{(i)}$ signals experiences an interference $I_i$ coming from the sensors that transmit through the same resources. This interference can be decomposed in what follows: (i) the interference coming from the sensors to be detected at the same beam and CN, namely intra-beam $I_i^{\textrm{intra}}$; and (ii) the interference coming from the sensors to be detected at the rest of beams and CNs that share the same resources, namely inter-beam $I_i^{\textrm{inter}}$. As~a result, we can~write
\begin{equation}
    I_i = I_{i}^{\textrm{intra}} + I_{i}^{\textrm{inter}},
\label{eq:27}
\end{equation}
where
\begin{equation}
    I_{i}^{\textrm{intra}} = \sum_{j \in \mathcal{J}_{k(i),l(i)} \backslash \{i\}} \bm{g}_{k(i),l(i)}^{\textrm{H}} \bm{h}_{k(i),j} \beta_j x_j , \quad
    I_{i}^{\textrm{inter}} = \sum_{\substack{(k,l) \neq (k(i),l(i)) \\
    t_i \in \mathcal{T}_{k,l}}} \sum_{j \in \mathcal{J}_{k,l}} \bm{g}_{k(i),l(i)}^{\textrm{H}} \bm{h}_{k(i),j} \beta_j x_j.
\label{eq:28}
\end{equation} 

The set $\mathcal{J}_{k,l}$ represents all the sensors detected at the CN--beam pair $(k,l)$, i.e.,~
\begin{equation}
    \mathcal{J}_{k,l} = \{j: (k(j),l(j)) = (k,l) \}.
\label{eq:29}
\end{equation}

Thereby, to~be consistent, we need to extract the signal from the $i$th sensor from the set $\mathcal{J}_{k(i),l(i)}$, as~shown in~Equation \eqref{eq:28}. The~term $t_i \in \{1,\ldots,R\}$ in~Equation \eqref{eq:28} denotes the identifier of the resource that the $i$th sensor is using, and~the set $\mathcal{T}_{k,l}$ contains the resources allocated to the CN--beam pair $(k,l)$. 

Overall, the~aggregated interference $\Gamma_i$ from~Equation \eqref{eq:2} is decomposed into the following:
\begin{equation}
    \Gamma_i = \Gamma_{i}^{\textrm{intra}} + \Gamma_{i}^{\textrm{inter}},
\label{eq:30}
\end{equation}
where the terms $\Gamma_{i}^{\textrm{intra}}$, $\Gamma_{i}^{\textrm{inter}}$ follow directly from~Equation \eqref{eq:28}:
\begin{equation}
    \Gamma_{i}^{\textrm{intra}} = \sum_{j \in \mathcal{J}_{k(i),l(i)} \backslash \{i\}} \beta_j a_{j,i}, \quad
    \Gamma_{i}^{\textrm{inter}} = \sum_{\substack{(k,l) \neq (k(i),l(i)) \\
    t_i \in \mathcal{T}_{k,l}}} \sum_{j \in \mathcal{J}_{k,l}} \beta_j a_{j,i}.
\label{eq:31}
\end{equation}

Similar to above, the~received power terms $a_{i,i}$ and $a_{j,i}$ read as~follows
\begin{equation}
    a_{i,i} = P \vert \bm{g}_{k(i),l(i)}^{\textrm{H}} \bm{h}_{k(i),i} \vert^2,
    \quad a_{j,i} = P \vert \bm{g}_{k(i),l(i)}^{\textrm{H}} \bm{h}_{k(i),j} \vert^2.
\label{eq:32}
\end{equation}

Now, the~interfering set $\mathcal{J}_i$ is:
\begin{equation}
    \mathcal{J}_i = \mathcal{J}_i^{\textrm{intra}} \cup \mathcal{J}_i^{\textrm{inter}},
\label{eq:33}
\end{equation}
where
\begin{equation}
    \mathcal{J}_i^{\textrm{intra}} = \mathcal{J}_{k(i),l(i)} \backslash \{i\}, \quad
    \mathcal{J}_i^{\textrm{inter}} = \bigcup_{\substack{(k,j) \neq (k(i),l(i)) \\
     t_i \in \mathcal{T}_{k,l}}} \mathcal{J}_{k,l} .
\label{eq:34}
\end{equation}

At this point, we can particularize $P_\textrm{out}^i$ from~Equation \eqref{eq:16} for the single- and multiple-resource~scenarios.

\subsubsection{Single-Resource~Scenario} 
In this case, only a single resource is allowed per CN--beam tuple. Accordingly, the~resource sets $\mathcal{T}_{k,l}$ have one element and sensor $i$ only uses a certain resource, denoted by $t_i$, which constitutes the set $\mathcal{T}_{k(i),l(i)}$, i.e.,~$\mathcal{T}_{k(i),l(i)} = \{t_i\}$. Hence, all sensors detected at the tuples $(k,l)$ with the same resource $t_i$, i.e.,~$\mathcal{T}_{k,j} = \{t_i\}$, create interference when detecting the signal coming from the $i$th sensor (cf.~Equation~\eqref{eq:34}).

As a result, $P_{\textrm{out}}^i$ from~Equation \eqref{eq:16} is completely defined by the pmf of the aggregated interference $\Gamma_i$, regardless of the resource $t_i$, which only determines the interfering set:
\begin{equation}
    P_{\textrm{out}}^i = \textrm{Pr}\{\Gamma_i > \xi_i - \sigma_{n,i}^2\} = \sum_{\gamma_i > \xi_i - \sigma_{n,i}^2} p_{\Gamma_i}(\gamma_i),
\label{eq:35}
\end{equation}
where we use $\xi_i = a_{i,i}/\delta$ for the sake of brevity in the notation. Note that, using the characteristic function method, a~numerical approximation of $P_{\textrm{out}}^i$ can be found. However, we use the continuous approximation $f_{\Gamma_i}^N(\gamma_i)$ of the pmf $p_{\Gamma_i}(\gamma_i)$ to express the outage probability from~Equation \eqref{eq:16} in closed-form:
\begin{equation}
    P_{\textrm{out}}^i = \sum_{\gamma_i > \xi_i - \sigma_{n,i}^2} p_{\Gamma_i}(\gamma_i) \approx \int_{\xi_i - \sigma_{n,i}^2}^{J_i} f_{\Gamma_i}^N(\gamma_i) d \gamma_i,
\label{eq:36}
\end{equation}
where $J_i$ refers to the upper bound of the finite support of $p_{\Gamma_i}(\gamma_i)$. This way, using~Equation \eqref{eq:36} together with the expression of $f_{\Gamma_i}^N(\gamma_i)$ derived in Section~\ref{sec:3.2}, we can obtain an analytic closed-form approximation for the expression of the outage probability $P_{\textrm{out}}^i$ defined in~Equation \eqref{eq:16}.

For instance, using $f_{\Gamma_i}^5(\gamma_i)$ from~Equation \eqref{eq:14}, i.e.,~set order to $N = 5$, the~approximation in~Equation \eqref{eq:36} yields
\begin{equation}
    P_{\textrm{out}}^i \approx \frac{1}{F_i} \int_{\xi_i - \sigma_{n,i}^2}^{J_i} \nu_i(\bar{\gamma}_i) \phi(\gamma_i;\mu_i,\sigma_i) d\gamma_i 
    = \frac{1}{F_i}\sum_{n = 0}^5 A_n \int_{\xi_i - \sigma_{n,i}^2}^{J_i} \bar{\gamma}_i^n \phi_s(\bar{\gamma}_i) d\bar{\gamma}_i = \frac{1}{F_i} \sum_{n = 0}^5 A_n^i G_n^i,
\label{eq:37}
\end{equation}
where we gather together the coefficients of the polynomials of equal order to get a more compact expression. The~integral terms $G_n^i$ can be found via Owen's T function~\cite{Owe80} and the terms $A_n^i$ are listed in the table below, where $B_n^i=B_n(\eta_1^i,\ldots,\eta_n^i)/(n ! \sigma_i^n)$ include the set of Bell polynomials: 

\begin{table}[H]
\caption{$A_n^i$ terms used in~Equation \eqref{eq:37}, listed from $n = 0$ to $n = 5$.}
\begin{center}
    {\renewcommand{\arraystretch}{1.2}
    \begin{tabular}{|c|c|c|c|c|c|c|} 
        \hline
        $n$ & $0$ & $1$ & $2$ & $3$ & $4$ & $5$\\
        \hline
        $A_n^i$ & $1 - B_2^i + 3 B_4^i$ & $B_1^i - 3 B_3^i + 15 B_5^i$ & $B_2^i - 6 B_4^i$ & $B_3^i - 10 B_5^i$ & $B_4^i$ & $B_5^i$\\ 
        \hline
    \end{tabular}}
\end{center}
\end{table}

It is important to highlight that this approximation allows us to work directly with the statistical moments of the aggregated interference instead of the instantaneous power values $a_{j,i}$. In~fact, the~cumulants needed for the Gram--Charlier series expansion are also obtained with these statistical moments. Therefore, the~outage probability defined in~Equation \eqref{eq:36} is completely characterized by those parameters together with the term $a_{i,i}$ (i.e., the received power of the signal from sensor $i$).

\subsubsection{Multiple-Resource~Scenario}
Equation \eqref{eq:36} is valid only when a single resource is allowed per CN--beam tuple. However, when considering that multiple resources can be allocated to each tuple, we need to generalize. Up~to now, $\beta_j$ has been a Bernoulli RV modeling the activity of sensor $j$. In~the multiple-resource case, $\beta_j$ is also a RV modeling the activity of the $j$th sensor when it is actually creating interference (cf.~Equation \eqref{eq:34}). Accordingly, for~the cases $t_i \in \mathcal{T}_{k(j),l(j)}$, $\beta_j$ can be decomposed as follows:
\begin{equation}
    \beta_j = \alpha_j \tau_j,
\label{eq:38}
\end{equation}
where $\alpha_j \sim$ \textit{Ber}($p_j^{\textrm{act}}$) represents the event of being active and transmitting. This RV depends only on the sensor itself and is equivalent to the RV in the single-resource setup. On~the other hand, $\tau_j$ is a Bernoulli RV, independent of $\alpha_j$, that is equal to $1$ whenever sensor $j$ selects randomly the same resource that is using the $i$th sensor, i.e.,~$t_i$. Then, in~the cases that $t_i \in \mathcal{T}_{k(j),l(j)}$, we have $\tau_j \sim$ \textit{Ber}($p_j^{\textrm{res}}$), where $p_j^{\textrm{res}} = 1/\vert \mathcal{T}_{k(j),l(j)} \vert $ assuming that sensors choose one of the resources within $\mathcal{T}_{k(j),l(j)}$ with equal probability. Hence, $\tau_j$ depends on the number of possible resources that sensor $j$ can equally choose, i.e.,~$1 \leq \vert \mathcal{T}_{k(j),l(j)} \vert \leq R$ and, thus, on~the tuple $(k(j),l(j))$ and the resource allocation. The case $\vert \mathcal{T}_{k(j),l(j)} \vert = 0$ is not considered since it corresponds to the situation when no resources are allocated to the tuple $(k(j),l(j))$. The~sensors detected at that pair are not included in the interfering set since $t_i \not\in \mathcal{T}_{k(j),l(j)}$ in that case. Besides, all RVs $\tau_j$ are assumed to be independent, including those from sensors detected at the same CN--beam tuple and that share the same parameter $1/\vert \mathcal{T}_{k(j),l(j)} \vert$.

As a result, given that both Bernoulli RVs are independent, $\beta_j$ is still a Bernoulli RV with parameter $p_j=p_j^{\textrm{act}}p_j^{\textrm{res}}$. 
In this case, $p_j$ represents the probability of being active and also of transmitting through the resource $t_i$ within the set $\mathcal{T}_{k(j),l(j)}$. Thus, the~use of multiple resources entails a reduction of sensors activity which, in~turn, reduces interference. In addition, note that, for the single-resource case, we only need to set $\tau_j = 1 \, \forall j$ or, equivalently, $\vert \mathcal{T}_{k(j),l(j)} \vert = 1 \, \forall j$. That is why the multiple-resource can be seen as a generalization of the single-resource~setup.

Furthermore, now interference may take place whenever there is non-null intersection between the resource sets, i.e.,~$\mathcal{T}_{k(i),l(i)} \cap \mathcal{T}_{k(j),l(j)} \neq \emptyset$. In~fact, since $t_i \in \mathcal{T}_{k(i),l(i)}$ might no longer be unique, the~interfering set $\mathcal{J}_i$ changes accordingly (cf.~Equation \eqref{eq:34}). Let us denote this sets by $\mathcal{J}_i(t_i)$ to include the dependence with the resource $t_i$. Therefore, although~the expression of the SINR in~Equation \eqref{eq:1} is still valid when the resource $t_i$ is used, we have $\vert \mathcal{T}_{k(i),l(i)} \vert$ different SINRs, one for each resource available for the $i$th sensor that is detected at the CN--beam tuple $(k(i),l(i))$.

Overall, since the reference sensor $i$ decides equally among $\vert \mathcal{T}_{k(i),l(i)} \vert$ resources, we need to include this random selection in the outage probability. To~this end, we average over the different possibilities, where the resource $t_i$ changes and so does the interfering set $\mathcal{J}_i(t_i)$:
\begin{equation}
    P_{\textrm{out}}^i = p_i^{\textrm{res}} \sum_{t_i \in \mathcal{T}_{k(i),l(i)}} \sum_{\gamma_i(t_i) > \xi_i - \sigma_{n,i}^2} p_{\Gamma_i(t_i)}(\gamma_i(t_i))
    = p_i^{\textrm{res}} \sum_{t_i \in \mathcal{T}_i} P_{\textrm{out}}^i (t_i),
\label{eq:39}
\end{equation}
where now the RV $\Gamma_i(t_i)$ also depends on the resource chosen by the $i$th sensor from the set of resources available at the CN--beam tuple $(k(i),l(i))$. Note that $P_{\textrm{out}}^i (t_i)$ is used to denote the outage probability when the resource $t_i \in \mathcal{T}_{k(i),l(i)}$ is used and can be approximated using~Equation \eqref{eq:36}. Thus, here we also end up with an analytic closed-form approximation for $P_{\textrm{out}}^i$ formulated in~Equation \eqref{eq:39}.

The previous analysis is of special interest in the next section, where we formulate and describe a possible solution for the resource allocation problem. Note that this task consists in deciding which resources from the set $\{1,\ldots,R\}$ are allocated to each CN--beam pair $(k,l)$. Accordingly, it ultimately defines the sets $\mathcal{T}_{k,l}$. To~do so, we use a graph-based approach that minimizes the average outage probability of all sensors within the~network.

\section{Resource Allocation~Problem} \label{sec:5}
Once the outage probability has been defined, we can design a strategy based on this magnitude to allocate the different resources in order to enhance the performance of the whole network. For~this task, in~the following, we present a graph-based approach that relies on the minimization of the previous approximation of the outage probability $P_{\textrm{out}}^i$. 

It is noteworthy to mention that the implementation of the proposed algorithm for the resource allocation is possible thanks to the analytic closed-form expression of the outage probability found in the previous section (which relies on the statistical moments of the aggregated interference).

Note that a good performance of the resulting resource allocation highlights the accuracy of the proposed Gram--Charlier approximation. Besides, given its relation to the sensors power consumption, lower values of $P_{\textrm{out}}^i$ result in a better energy efficiency, which is crucial in~mMTC.

As mentioned, we distinguish two scenarios: \textit{single}- and \textit{multiple-resource} scenarios. In~both setups, we~ seek for resource allocations that aim to minimize $P_{\textrm{out}}^i$. In~particular, we adopt a strategy in which the average network performance is optimized. However, any other approach could be used, e.g.,~a fair strategy where a certain minimum QoS is satisfied for all~sensors.

\begin{figure}[t]
    \centerline{\includegraphics[scale = 0.75]{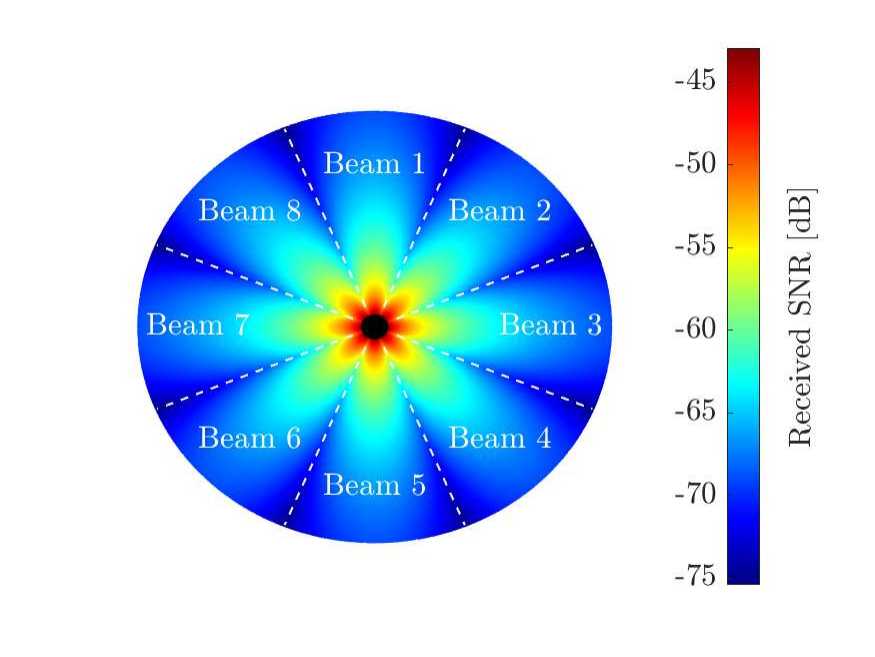}}
    \caption[]{Received SNR and spatial beams. A~minimum distance (black area) is set to avoid unrealistic high SNRs.}
\label{fig:2}
\end{figure}

Given that the positions of sensors are assumed to be fixed, the~CN--beam tuples where the sensors are detected are known. Recall that they refer to the pair with the highest received SNR according to~Equation \eqref{eq:25}. To~illustrate these tuples, let us consider a CN equipped with a uniform circular array (UCA). Note that this specific configuration is used here as an example, but~any other structure could be used. In addition, to~better understand the graphical representation, we assume the channels to be the steering vectors computed with the angle-of-arrival (AoA) of the sensors signals. A~simple set of spatial beam filters is then constructed with equispaced pointing directions. It is important to highlight that, for~a different array configuration and beamforming scheme, the~shape of the beams (i.e., the radiation pattern) changes and so do the portions in which the whole space is divided. This is ultimately represented by the set of tuples and, thus, once they are defined, the~following formulation is valid and the resource allocation mechanism remains the same.  In Figure~\ref{fig:2}, the received SNR at the beams of the CN is depicted for $N = L = 8$, unitary transmit and noise powers, and~a free-space path loss model. Sensors are deployed uniformly in a circle of radius $100$ m centered at the location of the~CN.

\subsection{Formulation in the Single and Multiple-Resource~Setups} \label{sec:5.1}
Let us start by defining an allocation matrix $\bm{C} \in \mathbb{N}^{K \times L}$ containing the resources allocated to each CN--beam tuple. Thereby, the~rows and columns represent the CNs and the beams, respectively. The~element $[\bm{C}]_{k,l}$ corresponds to the resources of the pair $(k,l)$.

In the \textit{single-resource} scenario, this matrix takes values within the set $\{0,\ldots,R\}$, where the zero refers to the case where no resource is allocated. Recall that here the sets $\mathcal{T}_{k,l}$ have a unique element. Hence, given that $\mathcal{T}_{k(i),l(i)} = \{t_i\}$, the~element $[\bm{C}]_{k(i),l(i)}$ is $t_i$. As~before, $t_i$ represents the identifier of the resource that the $i$th sensor is~employing.

We can relate the allocation matrix $\bm{C}$ to the interfering sets $\mathcal{J}_i$ in the following way:
\begin{equation}
    \mathcal{J}_i = \{j: [\bm{C}]_{k(j),l(j)} = t_i \} \backslash \{i\}.
\label{eq:40}
\end{equation}
 
On the other hand, a~different notation must be used in the \textit{multiple-resource} scenario. Now, the sets $\mathcal{T}_{k,l}$ can have more than one element. To represent all the possible combinations of resources, the~elements in the matrix $\bm{C}$ take values between $0$ and $2^R - 1$. Each element $[\bm{C}]_{k,l}$ corresponds to the decimal value of the binary vector $\bm{c}_{k,l} \in \{0,1\}^R$, i.e.,~
\begin{equation}
    [\bm{C}]_{k,l} = \sum_{n = 1}^R 2^n [\bm{c}_{k,l}]_n .
\label{eq:41} 
\end{equation}

Thereby, the~specific resources allocated to the tuple $(k,l)$ are indicated by the positions of the nonzero elements of the vector $\bm{c}_{k,l}$. For~instance, for~$R = 6$ and $\bm{c}_{k,l} = [1 0 1 0 0 1]$, the~element $[\bm{C}]_{k,l}$ would be $37$. In~that case, we would use the first, third, and sixth resource, i.e.,~$\mathcal{T}_{k,l} = \{1,3,6\}$. Note that the all zero vector is also allowed as the solution might ``switch off" completely some CN--beam tuples by no allocating resources to~them. 

Regarding the interfering sets $\mathcal{J}_i$, we have~that
\begin{equation}
    \mathcal{J}_i = \bigcup_{t_i \in \mathcal{T}_{k(i),l(i)}} \{j: [\bm{C}]_{k(j),l(j)} \in \mathcal{R}_{t_i} \} \backslash \{i\} ,
\label{eq:42} 
\end{equation}
where $\mathcal{R}_{t_i}$ contains all the elements in the set $\{0,\ldots,2^R - 1\}$ that include the usage of the resource $t_i$,~i.e.,
\begin{equation}
    \mathcal{R}_{t_i} = \{ r : r = \sum_{n = 1}^R 2^n [\bm{c}(t_i)]_n\},
\label{eq:43} 
\end{equation}
with $\{\bm{c}(t_i)\}$ being the set of binary vectors for which resource $t_i$ is being used, i.e.,~$[\bm{c}(t_i)]_{t_i} = 1$. For~instance, for~$R = 3$ and $t_i = 1$, these vectors would be $\{\bm{c}(1)\} = \{[100],[110],[101],[111]\}$ and, thus, the~corresponding set $\mathcal{R}_1$ would be $\{1,3,5,7\}$. 

Finally, since the purpose of this paper is to optimize the overall performance of the network, we look for allocation strategies that minimize the average outage probability of all the sensors. Consequently, the~resource allocation can be formulated as the following optimization problem:
\begin{equation}
    \bm{C}^{\star} = \underset{\bm{C}}{\mathrm{argmin}} \, \frac{1}{M} \sum_{i=1}^M P_{\textrm{out}}^i = \underset{\bm{C}}{\mathrm{argmin}} \, \bar{P}_{\textrm{out}},
\label{eq:44}
\end{equation}
where $P_{\textrm{out}}^i$ follows the definition in~Equations \eqref{eq:36} and~ \eqref{eq:39} for the single- and multiple-resource cases, respectively. Given the previous definitions, the~outage probability is completely defined by the resource allocation matrix $\bm{C}$. Note that, even though our approach has been formulated based on the objective function defined in~Equation \eqref{eq:44}, any other objective function could have been considered. For~instance, we could have used the maximum $P_{\textrm{out}}^i$, i.e.,
\begin{equation}
    \bm{C}^{\star} = \underset{\bm{C}}{\mathrm{argmin}} \, \underset{i}{\mathrm{max}} \, P_{\textrm{out}}^i.
\label{eq:45}
\end{equation}

It is straightforward to see that the problem formulated in~Equation \eqref{eq:44} is combinatorial and that the optimal solution has an exponential complexity: $\mathcal{O}((R + 1)^{KN})$ and $\mathcal{O}(2^{RKN})$ for the single- and multiple-resource cases, respectively. Hence, a~brute force approach is not affordable as trying all possibilities becomes quickly unfeasible. That is why we need to seek suboptimal strategies that provide a proper solution with far lower complexity. As~mentioned above, in~the forthcoming subsection, we present a graph-based approach for this task, which uses coloring techniques to achieve a feasible allocation given that we have a limited number of~resources.

\subsection{Solution Based on Graph~Coloring} \label{sec:5.2}
One way to solve the problem stated in~Equation \eqref{eq:44} is by means of graph coloring methods. In~this work, we present an approach that relies on graph structures that capture the previous setup reliably. We~ denote the CN--beam tuples as the nodes or vertices, and~an edge or connection is established whenever two pairs can potentially interfere (i.e., whenever the sensors signals detected at a pair can potentially interfere the other pair). Accordingly, we use geometrical measures to determine whether the interference is large enough to create a~connection. 

The whole graph can be represented with an adjacency matrix $\bm{A} \in \{0,1\}^{D \times D}$ \cite{Bal12}, where $D = LK$ is the number of nodes. Each entry $[\bm{A}]_{n,m}$ is $1$ for connecting nodes $n$ and $m$.

Given that CNs might be deployed at the center of areas where the sensor concentration is high, we define a circle of radius $R_\textrm{dep}$ around each CN to represent these regions. Then, with~the help of an interference radius $R_{\textrm{int}} = c R_{\textrm{dep}}$, we decide which of the interference coming from the sensors UL signals might be significant. The~factor $c \geq 1$ 
essentially determines how far the interfering sensors should be to be considered negligible (e.g., when the power of their received signal is orders of magnitude lower than that of sensor $i$).

Thereby, whenever the distance between two CNs is smaller than $R_{\textrm{int}} + R_{\textrm{dep}}$, we label them as potentially interfering CNs~\cite{Lag16}. Note that this is the condition for intersection between the circle describing the deployment area of sensors around one CN ($R_{\textrm{dep}}$) and the circle representing the range of the interference coming from the sensors deployed around another CN ($R_{\textrm{int}}$).  In addition, we denote their beams as interfering if one of them is pointing towards the other. Besides, all beams in the same CN are connected to each other. The~reason is that sensors detected by a CN--beam pair are probably close to that CN. Thus, it is likely that they create a large interference to the other beams in that CN (due to the secondary lobes of the beam radiation pattern). This procedure determines $\bm{A}$.

\begin{figure}[t]
    \centering
	\centerline{\includegraphics[scale = 0.75]{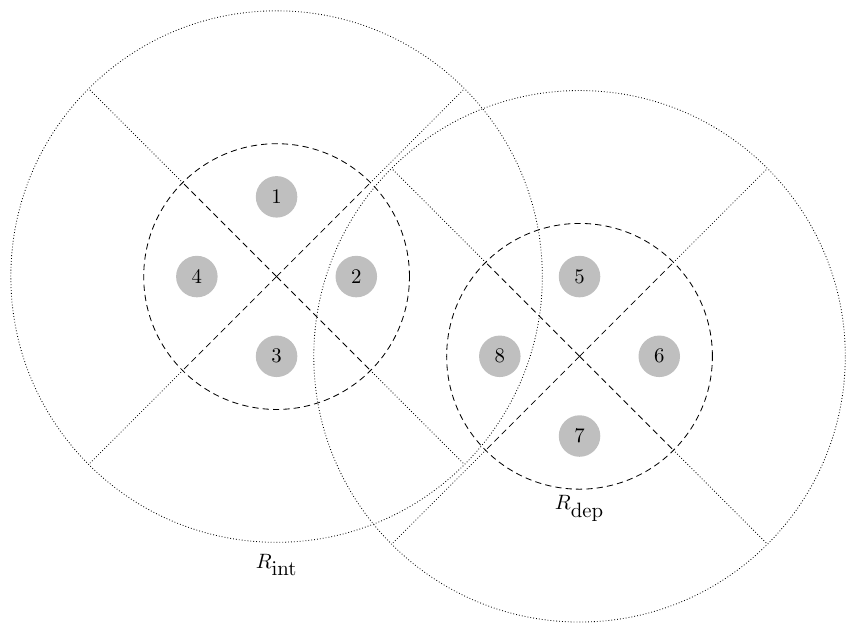}}
	\caption{Interfering nodes, listed from $1$ to $D = 8$.}
\label{fig:3}
\end{figure}

\begin{figure}[t]
    \centering
	\centerline{\includegraphics[scale = 0.75]{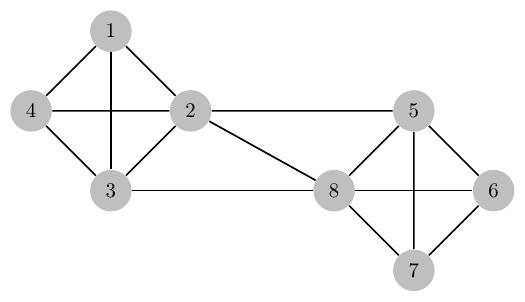}}
	\caption{Graph with resulting~connections.}
\label{fig:4}
\end{figure}

An example of the previous procedure is shown in Figures~\ref{fig:3} and \ref{fig:4} for $L = N = 4$, $K = 2$ and $c = 2$. In~addition, we consider a simple beamforming where the spatial filters are constructed with the steering vectors computed at equispaced pointing directions. However, the~following approach does not depend on the filtering scheme. From~now on, we consider that CNs use a generic beamforming. For~a different scheme, a~similar procedure can be used to find $\bm{A}$.

Once the graph is created, we color it with $R$ resources. In~the single-resource case, the~colors are directly the resource identifier. On~the contrary, in~the multiple-resource, they refer to the decimal value from~Equation \eqref{eq:41} that represents the set of resources of each pair. As~a result, we have $R + 1$ and $2^R$ colors for each scenario, respectively. In~both cases, we try to find an allocation such that two neighbors, i.e.,~connected nodes, do not share~resources.

In the single-resource scenario, having different resources at neighboring nodes is equivalent to having different colors. However, in~the multiple-resource, not only we need different colors, but~also we need to minimize the number of resources in common that they represent. For~example, for~$R = 6$ and colors $37$ and $53$, the~resource sets are $\{1,3,6\}$ and $\{1,3,5,6\}$, respectively. As~a result, even though the colors are different, the~resource sets have a non-null~intersection.

That is why we need to seek strategies that avoid any reuse of resources between neighbors (and~not only colors). In~the single-resource case, this is known as proper graph coloring~\cite{Bal12}. However, in~the multiple-resource, standard proper coloring does not guarantee that there is no reuse (i.e., null intersections between resource sets of connected nodes). Given the limited number of resources (e.g., in LTE-M, only $R = 6$ resources are destined to MTC~\cite{Nok16}), we allow two neighbors to share resources~\cite{Cow97}. Consequently, we must introduce some criterion to choose which nodes (i.e., CN--beam tuples) can reuse~resources. 

The proposed solution for the optimization problem in~Equation \eqref{eq:44} is the result of an iterative algorithm similar to the well-known first-fit (or greedy) approach used for graph coloring~\cite{Gro12}. To~ease of notation, 
we use the vectorized form of the allocation matrix $\bm{C}$, i.e.,~$\bm{c} = \textrm{vec} (\bm{C})$. 

The first step is to order the nodes according to the number of neighbors $\bm{n} = [n_1,\ldots,n_D]$ in a descent way, where $d\in \{1,\ldots,D\}$ is the node index. This magnitude is usually referred to as degree and captures roughly the amount of interference that the nodes can suffer. It can be computed as $\bm{n} = \bm{A}\textrm{diag}(\bm{I}_D)$. This ordering is represented with the vector $\bm{o} = [o_1,\ldots,o_D]$.

Next, we allocate random colors to each node. Afterwards, following the order given by $\bm{o}$, we~iterate over the nodes by assigning to each of them the color that leads to the minimum average outage probability. Note that this metric takes into account the non-null intersection of resources between different colors. This way, we follow the criterion in~Equation \eqref{eq:44}, and~the $o_d$th element of $\bm{c}$, i.e.,~$\bm{c}(o_d)$, is updated as:
\begin{equation}
    \mathcal{P}(W,o_d) = \underset{\bm{c}(o_d) \in \{0,\ldots,W\}}{\textrm{argmin}} \, \bar{P}_{\textrm{out}} (\bm{c}),
\label{eq:46}
\end{equation}
where $W$ is $R$ and $2^R - 1$ in the single- and multiple-resource case, respectively. At~each iteration, only one of the elements of vector $\bm{c}$ (that denoted by $\bm{c}(o_d)$) is allowed to change. The~procedure is summarized in Algorithm~\ref{alg:1}, where $\bar{P}_{\textrm{out}} ^{(u)}$ is the average outage probability at the $u$th iteration. Note that we always follow the direction of $\bar{P}_{\textrm{out}}$ decrease and that the routine terminates when the decrease becomes smaller than a threshold $\chi$ or when a number of iterations $U$ is~exceeded.

\begin{algorithm}[t] 
\caption{Greedy~optimization.} 
\label{alg:1} 
\begin{algorithmic}
    \State \( \bm{n} = [n_1,\ldots,n_{D}] = \bm{A} \textrm{diag}(\mathbf{I}_D) \) \Comment{Compute number of neighbors}
    \State \( \bm{o} = [o_1,\ldots,o_{D}] = \textrm{sort}(\bm{n}) \) \Comment{Order by highest degree}
    \State \( \bm{c} \gets \textrm{random} \) \Comment{Initialize color vector}
    \State \(u = 0\) \Comment{Initialize iteration counter}
    \State \( \bar{P}_{\textrm{out}}^{(u)} \gets \sum P_{\textrm{out}}^i (\bm{c}) / M\) \Comment{Initialize average outage probability}
    	\Repeat
    		\State \( u \gets u + 1 \) \Comment{Update iteration counter} 
    		\For{$d = 1:D$} \Comment{Iterate over nodes} 
    		\State \( \bm{c}(o_d) \gets \mathcal{P}(W,o_d) \) \Comment{Compute color as indicated in~Equation \eqref{eq:46}}
    		\EndFor
    		\State \( \bar{P}_{\textrm{out}}^{(u)} \gets \sum P_{\textrm{out}}^i (\bm{c}) / M\) \Comment{Update average outage probability}
    	\Until{$\bar{P}_{\textrm{out}}^{(u - 1)} - \bar{P}_{\textrm{out}}^{(u)} <\chi$ or $u > U$ }	\Comment{Stopping criteria}
\end{algorithmic}
\end{algorithm}

\section{Numerical~Results} \label{sec:6}
This section is devoted to present several numerical results to validate the approximation of the aggregated interference statistics introduced in Section~\ref{sec:3.2} and, thus, to~sustain the adequacy of this tool for calculating the outage probability derived in Section~\ref{sec:4.3}. Later, simulations to evaluate the performance of the allocation strategy described in Section~\ref{sec:5.2} are~shown.

In particular, we compare our approximation with the experimental results obtained with the characteristic function method from Section~\ref{sec:3.1}. Regarding the resource allocation, we compare it w.r.t. a random allocation in order to highlight the performance of our approach. For~both studies, we~ consider the system model from Section~\ref{sec:4.3}.

On the other hand, as~mentioned in Section~\ref{sec:4.3}, here we present a practical implementation of the mechanism that informs the sensors about the resources they can use. In addition, to~faithfully represent a realistic scenario, we use parameters and guidelines specified by 3GPP and ITU standards. That is why we dedicate an initial subsection to discuss all these~issues.

\subsection{Practical~Issues}
In LTE/LTE-A~\cite{Lie11,Gho11}, the~smallest resource unit is the physical resource block (PRB). It~corresponds to 
a time--frequency orthogonal resource that occupies a $0.5$ ms slot and a $180$ kHz bandwidth \cite{3GPP45820}. To~include the coexistence of MTC systems in cellular communications in our study, we adopt the frame structure specified by that~standard. 

As mentioned, we assume we dispose of $R$ available PRBs, which are allocated to the different CN--beam tuples through the graph-based approach described in Section~\ref{sec:5.2}. However, the~process through which sensors know the resources they can use has not been specified~yet.

Given that typical RACH based approaches are not suited for mMTC systems, we propose a methodology similar to that described in~\cite{Kel17}, where the resource identifiers are distributed among the spatial beams. Once resources are allocated to the beams, they are broadcast resource grants. Sensors detecting a certain grant, which means they are located in the beams pointing directions, use the associated PRBs to communicate. Recall that, in the multiple-resource scenario, sensors choose one of the available PRBs at random. In~the event of receiving PRBs grants from more than one beam, sensors may choose that with the highest SNR (assuming normalized power per beam), i.e.,~that from the CN--beam tuple defined in~Equation \eqref{eq:25}.

On the other hand, we consider that sensors are deployed uniformly in a circle of radius $R_{\textrm{dep}}$ centered at the CN location, and~that $M_k$ is approximately equal for all CNs. The~CNs are also uniformly distributed in a square area of side $R_t$, which is set to $1$ km to represent the typical dimensions of a LTE macrocell. In~turn, $R_{\textrm{dep}} = 100$ m to match those of a~microcell. 

To represent the multi-antenna technology used at the CNs, we use an UCA configuration~\cite{Van04}. In~addition to that, given the low mobility of sensors~\cite{Guo17,Alc18}, we consider channels to change very slowly and, thus, constant and known during the data transmission. For~simplicity, we assume that we have a Line-of-Sight communication. A possible extension to this work could be to analyze the case where the channel varies, incorporating fading in the communication, and/or where the channel is not perfectly known. 

Considering free-space propagation, the~sensors UL channels are expressed using the steering vector and the path-loss coefficient. For~the sake of simplicity, we consider the previous simple beamforming with $L = N$, where the spatial filters are constructed with the steering vectors computed at equispaced pointing directions. The~factor $c$ used to generate all the graphs is set to $2$ since the interfering signals coming from distances $d_j \geq R_{\textrm{int}} = 2 R_{\textrm{dep}}$ are received with a sufficiently high attenuation to be considered~negligible. 

Finally, the~probability $p_j^{\textrm{act}}$ is assumed the same for all sensors and equal to an activity factor $p$ (not~to be confused with the transmit power $P$). Nonetheless, in~the multiple-resource case, the~probabilities $p_j^{\textrm{res}}$ can be different as they depend on the number of resources allocated to the corresponding CN--beam tuple $(k(j),l(j))$. The~set of simulation parameters are listed in Table~\ref{tab:2}.

\begin{table}[H]
\caption{Simulation Parameters. The transmit power $P$ and the number of PRBs $R$ are selected following the LTE-M standard~\cite{Nok16}. According to the 3GPP indications in~\cite{3GPP45820}, low order constellations are used. For~a QPSK modulation, a~SINR of $-6.7$ dB is needed to achieve standard block error probabilities less than $10$ \% \cite{Gho11}. The~rest of parameters (e.g., carrier frequency) can be found in~\cite{ITU09}.}
\begin{center}
    \begin{tabular}{|c|c|c|c|c|c|c|c|}
        \hline
        $M$ & $K$& $N$ & $P$ & $p$& $R$ & $\delta$ \\ 
        \hline
        $2000$ & $10$ & $10$ & $0.1$ W & 0.1 & 6 & $-6.7$ dB \\
        \hline
    \end{tabular}
\end{center}
\label{tab:2}
\end{table}

\subsection{Aggregated Interference Statistics~Approximations}
\label{sec:6.1}
To show the accuracy of the Gram--Charlier approximation, we start by plotting the resulting cdf together with that obtained with the characteristic function. Recall that, for a sufficient number of points in the IFFT, the~method of the characteristic function represents a more precise approximation. That is why, since the real pmf of $\Gamma_i$ is not available, this approach is used as reference. All plots are done for a certain sensor $i$ at a random location. Note that only the single-resource case is shown as it is a sufficient representation of the setup. In comparison with the single-resource case, in~the multiple-resource case only the probabilities $p_j$ change. However, since the proposed Gram--Charlier approximation already takes into account that situation (i.e., is valid for any value of the probabilities $p_j$), the~shape of the resulting cdf in the multiple-resource does not differ from that of the single-resource. Thus, the~multiple-resource case is omitted to avoid redundancy. 

\begin{figure}[t]
    \centering
	\centerline{\includegraphics[scale = 0.75]{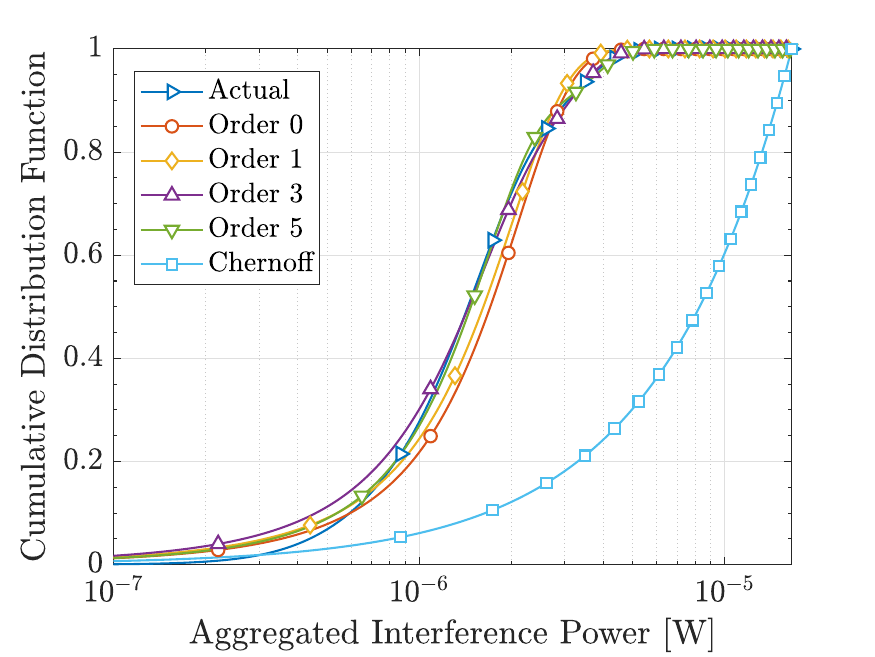}}
	\caption{ Actual and approximated~cdfs.} 
	\label{fig:5}
\end{figure}

Results are shown in Figure~\ref{fig:5}, where we can observe the accuracy of our approximation. Orders up to $5$ are presented to illustrate that the Gram--Charlier series expansion converges towards the actual pmf when using more addends. Note that order $0$ corresponds to the approach described in~\cite{SLMGS18}, where the Gaussian kernel is employed without any expansion. Besides, the~cdf of the Chernoff-based approximation from~\cite{SLMGS18} is included to make a broader comparison. As~expected, and~following the discussion in~\cite{SLMGS18}, it yields a large error. In~turn, our approach reveals an accurate approximation and a promising performance, specially for high~orders.

On the other hand, to~further illustrate the accuracy of our proposal, we analyze the error between the distribution obtained with the characteristic function, i.e.,~$p_{\Gamma_i}(\gamma_i)$, and~that provided by the Gram--Charlier (and Chernoff) method, i.e.,~$f_{\Gamma_i}^N(\gamma_i)$. To~compare them, we use the Jensen--Shannon divergence, which is a true distance and is bounded between $0$ and $1$ \cite{Lin91}:
\begin{equation}
\textrm{JSD}(p_{\Gamma_i}(\gamma_i),f_{\Gamma_i}^N(\gamma_i)) = \textrm{KL}(p_{\Gamma_i}(\gamma_i),m_{\Gamma_i}(\gamma_i))/2 + \textrm{KL}(f_{\Gamma_i}^N(\gamma_i),m_{\Gamma_i}(\gamma_i))/2,
\label{eq:47}
\end{equation}
where $m_{\Gamma_i}(\gamma_i) = (p_{\Gamma_i}(\gamma_i) + f_{\Gamma_i}^N(\gamma_i))/2$ represents the average distribution of $p_{\Gamma_i}(\gamma_i)$ and $f_{\Gamma_i}^N(\gamma_i)$, and~$\textrm{KL}(\cdot,\cdot)$ is the standard Kullback--Leibler divergence~\cite{Lin91}.

Thereby, we can define the error in our approximation~as
\begin{equation}
\varepsilon = \frac{1}{M} \sum_{i=1}^M \textrm{JSD}(p_{\Gamma_i}(\gamma_i),f_{\Gamma_i}^N(\gamma_i)),
\label{eq:48}
\end{equation}
which represents the average among all sensors. Note that, to compute the error $\varepsilon$ numerically, the~continuous approximations (Gram--Charlier and Chernoff) must be discretized. This is not necessary in the case of the characteristic function method since it already provides a~pmf.

\begin{figure}[t]
    \centering
	\centerline{\includegraphics[scale = 0.75]{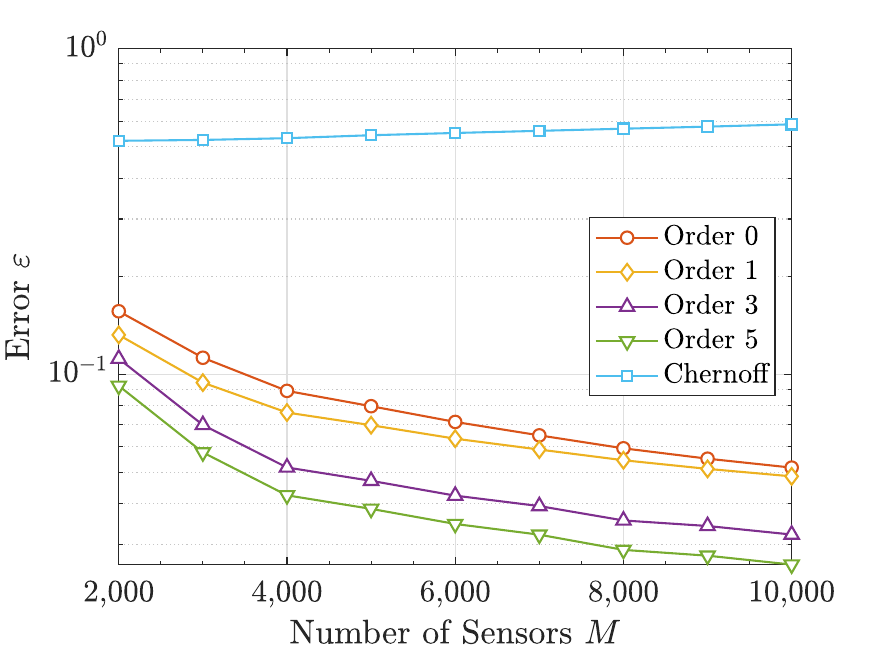}}
	\caption{ Error $\varepsilon$ versus number of sensors $M$.}
	\label{fig:6}
\end{figure}

\begin{figure}[t]
    \centering
	\centerline{\includegraphics[scale = 0.75]{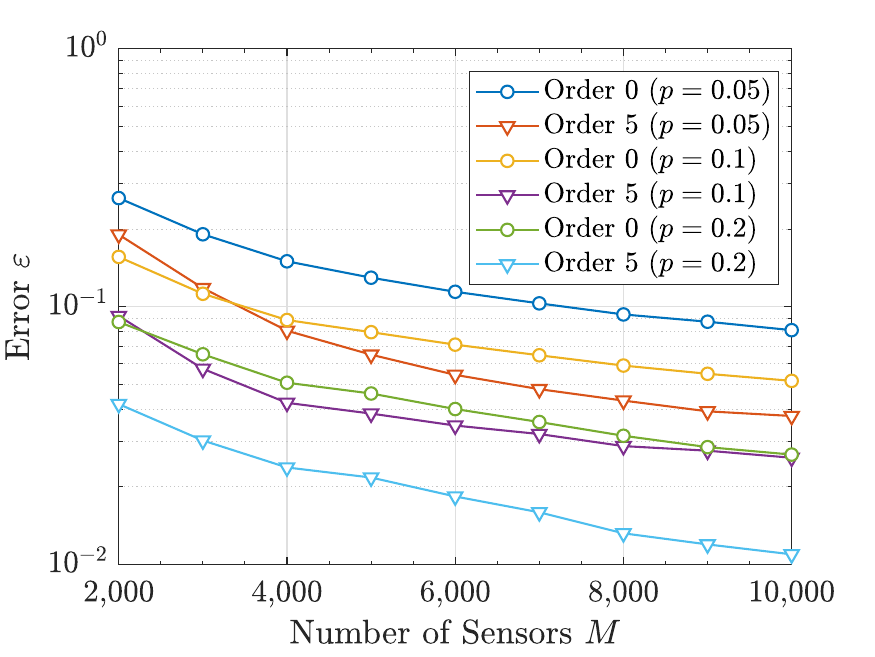}}
	\caption{Error $\varepsilon$ versus $M$ for different activity factors $p$.}
	\label{fig:7}
\end{figure}

Following the previous discussion, we compare the different approaches for the the single-resource scenario only. In~particular, we present the error in~Equation \eqref{eq:48} that each approach attains w.r.t. the number of sensors $M$. This is shown in Figures~\ref{fig:6} and \ref{fig:7}. In~the latter, the~error $\varepsilon$ is also depicted for different values of the activity factor $p$. This way, we can highlight the robustness of our proposal against the probability $p$. To~avoid redundancy, in Figure~\ref{fig:7} , only orders $0$ and $5$ are~shown.

It can be observed in Figure~\ref{fig:6} that $\varepsilon$ diminishes with the number of sensors $M$. This is due to the asymptotic behavior of the sum of RVs (CLT), i.e.,~the more addends the aggregated interference has (which is the correct assumption in mMTC), the~better the Gram--Charlier approximation becomes. Besides, the~error also decreases with $p$, as~shown in Figure~\ref{fig:7}. The~reason behind is that larger activities can be seen as an increase of the number of sensors creating interference, which makes the actual statistics to be closer to the asymptotic behavior. As~before, a~finer precision is attained for higher orders and a poor performance is obtained with the Chernoff~method.

\subsection{Resource~Allocation} \label{sec:6.2}
In this section, we present different results to assess the performance of the allocation strategy described in Section~\ref{sec:5.2}. These simulations are used to illustrate the enhancement w.r.t. the trivial allocation where the entries of $\bm{C}$ are selected randomly between 0 and $W$.

\begin{figure}[t]
    \centering
	\centerline{\includegraphics[scale = 0.75]{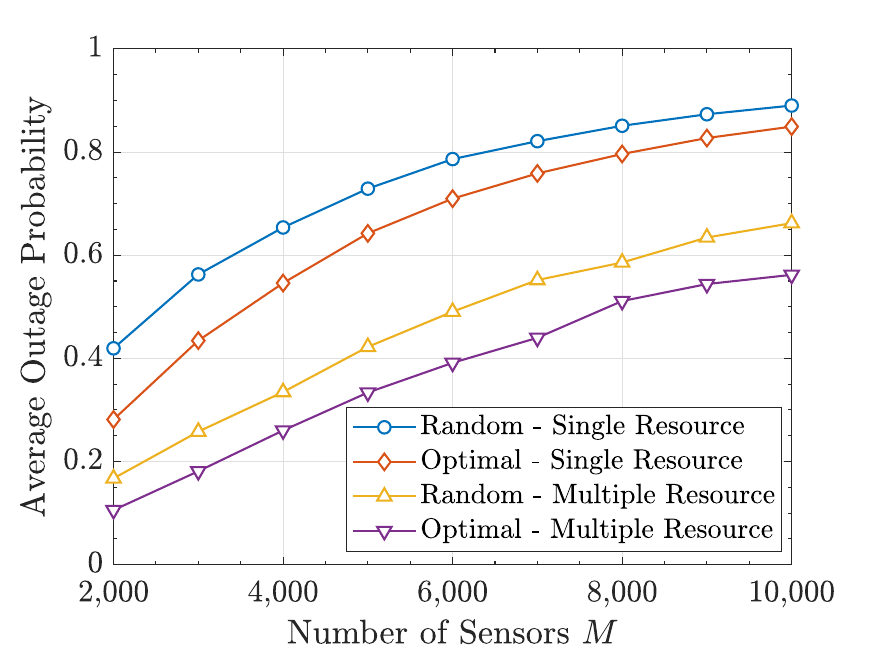}}
	\caption{Average outage probability $\bar{P}_{\textrm{out}}$ versus $M$.}
	\label{fig:8}
\end{figure}

\begin{figure}[t]
    \centering
	\centerline{\includegraphics[scale = 0.75]{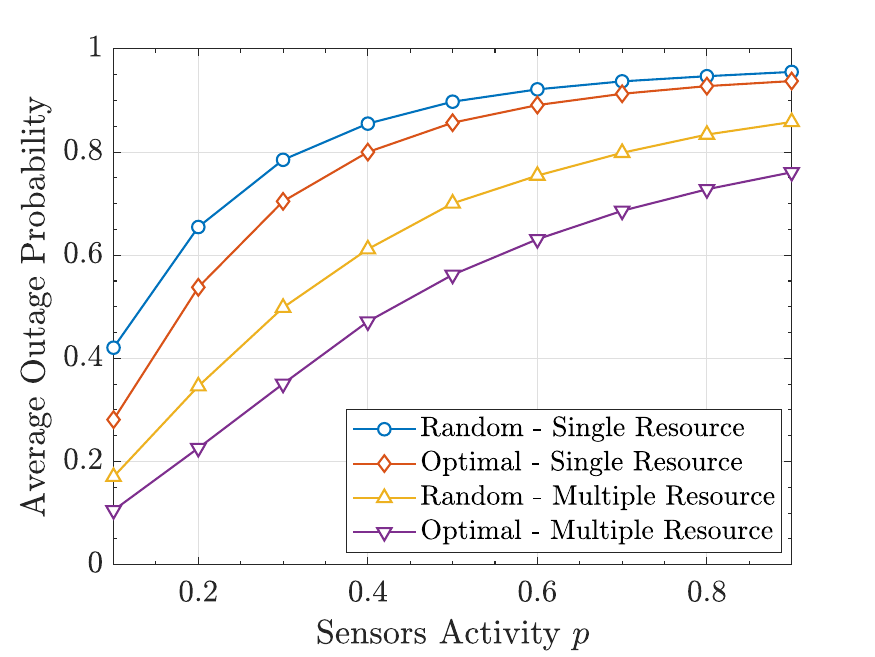}}
	\caption{Average outage probability $\bar{P}_{\textrm{out}}$ versus $p$.}
	\label{fig:9}
\end{figure}

We first plot the average outage probability $\bar{P}_{\textrm{out}}$ from~Equation \eqref{eq:44} obtained with both allocation strategies when changing $M$ and $p$. This is depicted in Figures~\ref{fig:8} and \ref{fig:9}, respectively. It can be seen that a substantial improvement is obtained with our proposal. In~turn, the~multiple-resource yields lower $\bar{P}_{\textrm{out}}$ as the activity is reduced and more degrees of freedom (i.e., colors) are available. Note that the resource allocation is always done with the Gram--Charlier approximation of order $5$ and that the resulting probability values are computed using the characteristic function method. 

\begin{figure}[t]
    \centering
	\centerline{\includegraphics[scale = 0.75]{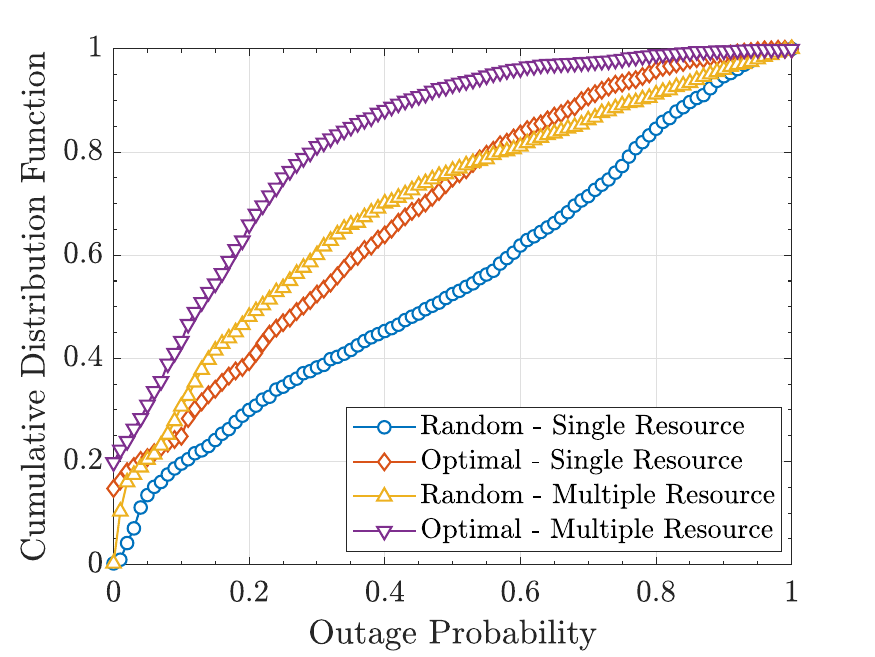}}
	\caption{CDF of the outage probability $F_{P_{\textrm{out}}^i} (p_{\textrm{out}}^i)$.}
	\label{fig:10}
\end{figure}

On the other hand, to~get richer insights, in~Figure~\ref{fig:10}, we show the cdf of the outage probability:
\begin{equation}
F_{P_{\textrm{out}}^i} (p_{\textrm{out}}^i) = \textrm{Pr}\{P_{\textrm{out}}^i \leq p_{\textrm{out}}^i\}.
\label{eq:49}
\end{equation}

As we can see, our strategy helps to decrease considerably the outage probabilities within the network. Therefore, given the relation between the outage probability and the consumed power, our~approach could also be useful to improve the energy efficiency of mMTC systems and extend the battery lifetime of~sensors.

\section{Conclusions} \label{sec:7}
In this paper, we address the problem of how to model the aggregated interference statistics, which captures the sporadic activity of sensors in the context of UL mMTC. Given its discrete nature and the large number of devices, the~expression of this distribution can be difficult to find. That is why we propose a Gram--Charlier series expansion of a truncated Gaussian kernel to approximate the aggregated interference statistics. Thanks to that, we derive an analytic closed-form expression for the outage probability, which is a valuable figure of~merit. 

In particular, we consider a scenario with several multi-antenna CNs that receive information from a group of sensors. Each of the CNs is equipped with a set of predefined spatial beams. We~distinguish two scenarios, single- or multiple-resource, depending on the number of resources allocated to each beam. Since the total number of resources is limited, we present a graph coloring technique that tries to minimize the average outage probability as allocation strategy. Finally, we~describe a practical mechanism where resource grants are sent through the beams in a broadcast way. Sensors located at their pointing directions will receive those permissions and use those resources to communicate. This non-dedicated scheduling approach can serve as an alternative to typical RACH schemes, which fail in the presence of massive~requests. 

Simulation results show that our proposal yields an accurate approximation and that our allocation method can improve the overall system~performance.

Based on the justifications detailed in the paper, the~random nature of the scenario comes from the sporadic activity and the high number of devices, while the channel is assumed to be known and fading is not included. Future work will focus on the extension to the cases where these assumptions do not hold. Note, however, that the performance predicted by the results in this paper can be taken as a valid benchmark for comparison purposes in those cases.

\bibliographystyle{IEEEtran}
\bibliography{IEEEabrv,References}

\begin{thebibliography}{10}
\providecommand{\url}[1]{#1}
\csname url@samestyle\endcsname
\providecommand{\newblock}{\relax}
\providecommand{\bibinfo}[2]{#2}
\providecommand{\BIBentrySTDinterwordspacing}{\spaceskip=0pt\relax}
\providecommand{\BIBentryALTinterwordstretchfactor}{4}
\providecommand{\BIBentryALTinterwordspacing}{\spaceskip=\fontdimen2\font plus
\BIBentryALTinterwordstretchfactor\fontdimen3\font minus \fontdimen4\font\relax}
\providecommand{\BIBforeignlanguage}[2]{{%
\expandafter\ifx\csname l@#1\endcsname\relax
\typeout{** WARNING: IEEEtran.bst: No hyphenation pattern has been}%
\typeout{** loaded for the language `#1'. Using the pattern for}%
\typeout{** the default language instead.}%
\else
\language=\csname l@#1\endcsname
\fi
#2}}
\providecommand{\BIBdecl}{\relax}
\BIBdecl

\bibitem{Wan17}
H.~Wang and A.~O. Fapojuwo, ``A survey of enabling technologies of low power and long range machine-to-machine communications,'' \emph{IEEE Commun. Surveys Tuts.}, vol.~19, no.~4, pp. 2621--2639, 4th Quart., 2017.

\bibitem{Sha15}
H.~Shariatmadari \emph{et~al.}, ``Machine-type communications: Current status and future perspectives toward {5G} systems,'' \emph{IEEE Commun. Mag.}, vol.~53, no.~9, pp. 10--17, Sep. 2015.

\bibitem{Xu18}
J.~Xu \emph{et~al.}, ``Narrowband {Internet of Things}: Evolutions, technologies, and open issues,'' \emph{IEEE Internet Things J.}, vol.~5, no.~3, pp. 1449--1462, Jun. 2018.

\bibitem{Teh14}
M.~N. Tehrani, M.~Uysal, and H.~Yanikomeroglu, ``Device-to-device communication in {5G} cellular networks: {Challenges}, solutions, and future directions,'' \emph{IEEE Commun. Mag.}, vol.~52, no.~5, pp. 86--92, May 2014.

\bibitem{And14}
J.~G. Andrews \emph{et~al.}, ``What will {5G} be?'' \emph{IEEE J. Sel. Areas Commun.}, vol.~32, no.~6, pp. 1065--1082, Jun. 2014.

\bibitem{Xi15}
X.~Xiong \emph{et~al.}, ``Low power wide area machine-to-machine networks: Key techniques and prototype,'' \emph{IEEE Commun. Mag.}, vol.~53, no.~9, pp. 64--71, Sep. 2015.

\bibitem{Agi16}
M.~Agiwal, A.~Roy, and N.~Saxena, ``Next generation {5G} wireless networks: A comprehensive survey,'' \emph{IEEE Commun. Surveys Tuts.}, vol.~18, no.~3, pp. 1617--1655, 3rd Quart., 2016.

\bibitem{Yan17}
W.~Yang \emph{et~al.}, ``Narrowband wireless access for low-power massive internet of things: A bandwidth perspective,'' \emph{IEEE Wireless Commun.}, vol.~24, no.~3, pp. 138--145, Jun. 2017.

\bibitem{Nok16}
{Nokia White Paper}, ``{LTE} evolution for {IoT} connectivity,'' 3rd Ed. 2016.

\bibitem{Els17}
M.~Elsaadany, A.~Ali, and W.~Hamouda, ``Cellular {LTE-A} technologies for the future {Internet-of-Things}: Physical layer features and challenges,'' \emph{IEEE Commun. Surveys Tuts.}, vol.~19, no.~4, pp. 2544--2572, 4th Quart., 2017.

\bibitem{Gaz17}
V.~Gazis, ``A survey of standards for machine-to-machine and the {Internet of Things},'' \emph{IEEE Commun. Surveys Tuts.}, vol.~19, no.~1, pp. 482--511, 1st Quart., 2017.

\bibitem{Daw17}
Z.~Dawy \emph{et~al.}, ``Toward massive machine type cellular communications,'' \emph{IEEE Wireless Commun.}, vol.~24, no.~1, pp. 120--128, Feb. 2017.

\bibitem{Ali15}
A.~Ali, W.~Hamouda, and M.~Uysal, ``Next generation {M2M} cellular networks: challenges and practical considerations,'' \emph{IEEE Commun. Mag.}, vol.~53, no.~9, pp. 18--24, Sep. 2015.

\bibitem{Boc16}
C.~Bockelmann \emph{et~al.}, ``Massive machine-type communications in {5G}: {Physical} and {MAC}-layer solutions,'' \emph{IEEE Commun. Mag.}, vol.~54, no.~9, pp. 59--65, Sep. 2016.

\bibitem{Eri16}
{Ericsson White Paper}, ``{5G} radio access,'' Apr. 2016.

\bibitem{Che17}
S.~Chen \emph{et~al.}, ``Machine-to-machine communications in ultra-dense networks - {A} survey,'' \emph{IEEE Commun. Surveys Tuts.}, vol.~19, no.~3, pp. 1478--1503, 3rd Quart., 2017.

\bibitem{3GPP45820}
{3GPP TR 45.820}, ``Cellular system support for ultra low complexity and low throughput {Internet of Things},'' Aug. 2016.

\bibitem{Tse05}
D.~Tse and P.~Viswanath, \emph{Fundamentals of Wireless Communication}.\hskip 1em plus 0.5em minus 0.4em\relax Cambridge University Press, 2005.

\bibitem{Lie11}
S.~Y. Lien, K.~C. Chen, and Y.~Lin, ``Toward ubiquitous massive accesses in {3GPP} machine-to-machine communications,'' \emph{IEEE Commun. Mag.}, vol.~49, no.~4, pp. 66--74, Apr. 2011.

\bibitem{Gha15}
F.~{Ghavimi} and H.~{Chen}, ``{M2M} communications in {3GPP LTE/LTE-A} networks: Architectures, service requirements, challenges, and applications,'' \emph{IEEE Commun. Surveys Tuts.}, vol.~17, no.~2, pp. 525--549, 2nd Quart., 2015.

\bibitem{Has13}
M.~Hasan, E.~Hossain, and D.~Niyato, ``Random access for machine-to-machine communication in {LTE-Advanced} networks: {Issues} and approaches,'' \emph{IEEE Commun. Mag.}, vol.~51, no.~6, pp. 86--93, Jun. 2013.

\bibitem{Kat00}
I.~Katzela and M.~Naghshineh, ``Channel assignment schemes for cellular mobile telecommunication systems: A comprehensive survey,'' \emph{IEEE Commun. Surveys Tuts.}, vol.~3, no.~2, pp. 10--31, 2nd Quart., 2000.

\bibitem{Rab11}
A.~{Rabbachin}, T.~Q.~S. {Quek}, H.~{Shin}, and M.~Z. {Win}, ``Cognitive network interference,'' \emph{IEEE J. Sel. Areas Commun.}, vol.~29, no.~2, pp. 480--493, Feb. 2011.

\bibitem{Kus12}
S.~{Kusaladharma} and C.~{Tellambura}, ``Aggregate interference analysis for underlay cognitive radio networks,'' \emph{IEEE Wireless Commun. Lett.}, vol.~1, no.~6, pp. 641--644, Dec. 2012.

\bibitem{R1}
G.~C. {Alexandropoulos}, P.~{Ferrand}, J.~{Gorce}, and C.~B. {Papadias}, ``Advanced coordinated beamforming for the downlink of future {LTE} cellular networks,'' \emph{IEEE Commun. Mag.}, vol.~54, no.~7, pp. 54--60, July 2016.

\bibitem{R2}
G.~C. {Alexandropoulos}, P.~{Ferrand}, and C.~B. {Papadias}, ``On the robustness of coordinated beamforming to uncoordinated interference and {CSI} uncertainty,'' in \emph{2017 IEEE Wireless Communications and Networking Conference (WCNC)}, March 2017, pp. 1--6.

\bibitem{R3}
R.~W. {Heath Jr}, T.~{Wu}, Y.~H. {Kwon}, and A.~C.~K. {Soong}, ``Multiuser {MIMO} in distributed antenna systems with out-of-cell interference,'' \emph{IEEE Trans. Signal Process.}, vol.~59, no.~10, pp. 4885--4899, Oct 2011.

\bibitem{Car10}
P.~{Cardieri}, ``Modeling interference in wireless ad hoc networks,'' \emph{IEEE Commun. Surveys Tuts.}, vol.~12, no.~4, pp. 551--572, 4th Quart., 2010.

\bibitem{ElS13}
H.~ElSawy, E.~Hossain, and M.~Haenggi, ``Stochastic geometry for modeling, analysis, and design of multi-tier and cognitive cellular wireless networks: {A} survey,'' \emph{IEEE Commun. Surveys Tuts.}, vol.~15, no.~3, pp. 996--1019, 3rd Quart., 2013.

\bibitem{Hae09}
M.~Haenggi \emph{et~al.}, ``Stochastic geometry and random graphs for the analysis and design of wireless networks,'' \emph{IEEE J. Sel. Areas Commun.}, vol.~27, no.~7, pp. 1029--1046, Sep. 2009.

\bibitem{And10}
J.~G. Andrews \emph{et~al.}, ``A primer on spatial modeling and analysis in wireless networks,'' \emph{IEEE Commun. Mag.}, vol.~48, no.~11, pp. 156--163, Nov. 2010.

\bibitem{Kwo13}
T.~Kwon and J.~M. Cioffi, ``Random deployment of data collectors for serving randomly-located sensors,'' \emph{IEEE Trans. Wireless Commun.}, vol.~12, no.~6, pp. 2556--2565, Jun. 2013.

\bibitem{Sen18}
K.~{Senel} and E.~G. {Larsson}, ``Grant-free massive {MTC}-enabled massive {MIMO: A} compressive sensing approach,'' \emph{IEEE Trans. Commun.}, vol.~66, no.~12, pp. 6164--6175, Dec. 2018.

\bibitem{Kel17}
P.~Kela \emph{et~al.}, ``Connectionless access for massive machine type communications in ultra-dense networks,'' in \emph{2017 IEEE International Conference on Communications}, May 2017, pp. 1--6.

\bibitem{Oh15}
C.~{Oh}, D.~{Hwang}, and T.~{Lee}, ``Joint access control and resource allocation for concurrent and massive access of {M2M} devices,'' \emph{IEEE Trans. Wireless Commun.}, vol.~14, no.~8, pp. 4182--4192, Aug 2015.

\bibitem{Si15}
P.~Si \emph{et~al.}, ``Adaptive massive access management for {QoS} guarantees in {M2M} communications,'' \emph{IEEE Trans. Veh. Technol.}, vol.~64, no.~7, pp. 3152--3166, Jul. 2015.

\bibitem{Lay14}
A.~Laya, L.~Alonso, and J.~Alonso-Zarate, ``Is the random access channel of {LTE} and {LTE-A} suitable for {M2M} communications? {A} survey of alternatives,'' \emph{IEEE Commun. Surveys Tuts.}, vol.~16, no.~1, pp. 4--16, 1st Quart., 2014.

\bibitem{Lag16}
S.~Lagen \emph{et~al.}, ``Long-term provisioning of radio resources based on their utilization in dense {OFDMA} networks,'' in \emph{2016 IEEE 27th Annual International Symposium on Personal, Indoor, and Mobile Radio Communications}, Sep. 2016, pp. 1--7.

\bibitem{Sal18}
T.~{Salam}, W.~U. {Rehman}, and X.~{Tao}, ``Cooperative data aggregation and dynamic resource allocation for massive machine type communication,'' \emph{IEEE Access}, vol.~6, pp. 4145--4158, 2018.

\bibitem{Alc18}
O.~L. {Alcaraz L\'opez} \emph{et~al.}, ``Aggregation and resource scheduling in machine-type communication networks: A stochastic geometry approach,'' \emph{IEEE Trans. Wireless Commun.}, vol.~17, no.~7, pp. 4750--4765, Jul. 2018.

\bibitem{Guo17}
J.~{Guo} \emph{et~al.}, ``Massive machine type communication with data aggregation and resource scheduling,'' \emph{IEEE Trans. Commun.}, vol.~65, no.~9, pp. 4012--4026, Sep. 2017.

\bibitem{Xia18}
N.~{Xia}, H.~{Chen}, and C.~{Yang}, ``Radio resource management in machine-to-machine communications—{A} survey,'' \emph{IEEE Commun. Surveys Tuts.}, vol.~20, no.~1, pp. 791--828, 1st Quart., 2018.

\bibitem{SLMGS18}
S.~{Liesegang}, O.~{Mu{\~{n}}oz}, and A.~{Pascual-Iserte}, ``Interference statistics approximations for data rate analysis in uplink massive {MTC},'' in \emph{2018 IEEE Global Conference on Signal and Information Processing}, Nov. 2018, pp. 176--180.

\bibitem{Bre17}
T.~Brenn and S.~N. Anfinsen, ``A revisit of the {Gram-Charlier} and {Edgeworth} series expansions,'' UiT The Arctic University of Norway, Department of Physics and Technology, Tech. Rep., Jun. 2017.

\bibitem{Com12}
L.~Comtet, \emph{Advanced Combinatorics: The Art of Finite and Infinite Expansions}.\hskip 1em plus 0.5em minus 0.4em\relax Springer Netherlands, 2012.

\bibitem{Smi95}
P.~J. Smith, ``A recursive formulation of the old problem of obtaining moments from cumulants and vice versa,'' \emph{The American Statistician}, vol.~49, no.~2, pp. 217--218, 1995.

\bibitem{Bur14}
J.~Burkardt, ``The truncated normal distribution,'' \emph{Dept. Scientific Computing Website, Florida State University}, Oct. 2014.

\bibitem{Zhe12}
K.~Zheng \emph{et~al.}, ``Radio resource allocation in {LTE-Advanced} cellular networks with {M2M} communications,'' \emph{IEEE Commun. Mag.}, vol.~50, no.~7, pp. 184--192, Jul. 2012.

\bibitem{Owe80}
D.~B. Owen, ``A table of normal integrals,'' \emph{Communications in Statistics - Simulation and Computation}, vol.~9, no.~4, pp. 389--419, 1980.

\bibitem{Bal12}
R.~Balakrishnan and K.~Ranganathan, \emph{A Textbook of Graph Theory}.\hskip 1em plus 0.5em minus 0.4em\relax Springer New York, 2012.

\bibitem{Cow97}
L.~Cowen, W.~Goddard, and C.~E. Jesurum, ``Defective coloring revisited,'' \emph{Journal of Graph Theory}, vol.~24, no.~3, pp. 205--219, 1997.

\bibitem{Gro12}
M.~Gr{\"o}tschel, L.~Lovasz, and A.~Schrijver, \emph{Geometric Algorithms and Combinatorial Optimization}.\hskip 1em plus 0.5em minus 0.4em\relax Springer Berlin Heidelberg, 2012.

\bibitem{Gho11}
A.~Ghosh and R.~Ratasuk, \emph{Essentials of LTE and LTE-A}.\hskip 1em plus 0.5em minus 0.4em\relax Cambridge University Press, 2011.

\bibitem{Van04}
H.~Van~Trees, \emph{Optimum Array Processing: Part IV of Detection, Estimation, and Modulation Theory}.\hskip 1em plus 0.5em minus 0.4em\relax Wiley, 2004.

\bibitem{ITU09}
ITU, ``Guidelines for evaluation of radio interface technologies for {IMT}-advanced,'' International Telecommunication Union (ITU), Tech. Rep., 2009, {ITU-R} (M.2135-1).

\bibitem{Lin91}
J.~{Lin}, ``Divergence measures based on the {Shannon} entropy,'' \emph{IEEE Trans. Inf. Theory}, vol.~37, no.~1, pp. 145--151, Jan. 1991.

\end{thebibliography}

\end{document}